\documentclass[twocolumn]{emulateapj}
\usepackage{apjfonts}

\usepackage{graphicx}
\usepackage{mathrsfs}
\usepackage{natbib}
\bibpunct[,]{(}{)}{;}{a}{}{,}

\def\totd{{\mathrm{d}}}
\def\rs0{{r_\mathrm{s0}}}
\def\vff2{{v_\mathrm{ff\,0}^2}}

\def\Xeq{{X_\alpha^{\rm eq}}}
\def\tff{{t_\mathrm{ff0}}}

\shorttitle{ALPHA PARTICLE RECOMBINATION IN SUPERNOVAE}
\shortauthors{FERN\'ANDEZ \& THOMPSON}
\slugcomment{Submitted 2008 December 30; accepted 2009 July 2; published 2009 September 10}

\begin{document}

\title{Dynamics of a Spherical Accretion Shock with \\ Neutrino Heating and Alpha-Particle Recombination }
\author{Rodrigo Fern\'andez\altaffilmark{1} and Christopher Thompson\altaffilmark{2}}
\altaffiltext{1}{Department of Astronomy and Astrophysics, University of Toronto. 
Toronto, Ontario M5S 3H4, Canada.}
\altaffiltext{2}{CITA, 60 St. George St., 
Toronto, Ontario M5S 3H8, Canada.}

\begin{abstract}
We investigate the effects of neutrino heating and $\alpha$-particle recombination 
on the hydrodynamics of core-collapse supernovae.  Our focus is on the 
non-linear dynamics of the shock wave that forms in the collapse, 
and the assembly of positive energy material below it.  To this end, we perform 
time-dependent hydrodynamic simulations with FLASH2.5 in spherical and axial symmetry.
These generalize our previous calculations by allowing for bulk neutrino heating
and for nuclear statistical equilibrium between $n$, $p$ and $\alpha$.  
The heating rate is freely tunable, as is the starting radius of the shock
relative to the recombination radius of $\alpha$-particles.  An explosion in 
spherical symmetry involves the excitation of an overstable mode, which may be viewed as the
$\ell = 0$ version of the `Standing Accretion Shock Instability'.  
In two-dimensional simulations, non-spherical deformations of the shock are driven by plumes of material
with positive Bernoulli parameter, which are concentrated well outside the zone
of strong neutrino heating.  The non-spherical modes of the shock reach a large
amplitude only when the heating rate is also high enough to excite convection below the shock.
The critical heating rate that causes
an explosion depends sensitively on the initial position of the shock
relative to the recombination radius.  Weaker heating is required to 
drive an explosion in two dimensions than in one, but the difference also depends
on the size of the shock.  Forcing the infalling heavy nuclei to break up
into $n$ and $p$ below the shock only causes a slight increase in the critical
heating rate, except when the shock starts out at a large radius.
This shows that heating by neutrinos (or some other mechanism)
must play a significant role in pushing the shock far enough out that
recombination heating takes over.
\end{abstract}


\maketitle

\section{Introduction}

Although tremendous progress has been made on the mechanism of
core-collapse supernovae in recent years, we still do not have a clear
picture of a robust path to an explosion in stars that form iron cores.  
Heating by the absorption
of electron-type neutrinos significantly modifies the settling flow
below the bounce shock but -- in spite of early positive results
\citep{bethe85} -- explosions are obtained 
in spherical collapse calculations 
only if the progenitor star is lighter than about 10-12 $M_\sun$
\citep{kitaura06}.  
More massive stars fail to explode in spherical 
symmetry \citep{liebendoerfer01}.

Two-dimensional collapse calculations show strong deformations
of the shock and convective motions below it
\citep{burrows95,janka96,buras06a,buras06b,burrows06a,burrows07,marek07}.
It has long been recognized that
convection increases the residency time of settling material in the zone of
strong neutrino heating \citep{herant92}.  It is also becoming clear that
multidimensional explosions require the assembly of a smaller amount of material
with positive energy, but the details of how this happens remain murky. 

An early treatment of shock breakout by \citet{bethe97} focused on
the strong gradient in the ram pressure of the infalling material, but implicitly
assumed that the shocked material had already gained positive energy.
If large-scale density inhomogeneities are present below the shock, they will
trigger a finite-amplitude, dipolar instability, thereby allowing accretion to
continue simultaneously with the expansion of positive-energy fluid (Thompson 2000).
The accretion shock is also capable of a dipolar oscillation which leads, above a
critical amplitude, to 
a bifurcation between freshly infalling material
and material shocked at earlier times \citep{BM03}. 
Such an oscillation
is easily excited in a spherical flow composed of a zero-energy, polytropic
fluid (\citealt{BM06}) via a linear feedback between ingoing vortex and entropy waves
and outgoing sound waves \citep{F07}.  It can also be excited
indirectly by neutrino heating, which if sufficiently strong will drive
large-scale buoyant motions below the shock 
(\citealt{herant94}; \citealt{foglizzo06}).  
For relatively weak heating, the dipolar oscillation can
decrease the damping effect of advection and trigger convective
motions that would otherwise be suppressed \citep{foglizzo06,scheck08}.

A significant sink of thermal energy in the accretion flow
arises from the dissociation of heavy nuclei. 
The heavy elements that flow through the shock
are broken up into $\alpha$-particles and nucleons 
when exposed to the high temperature 
($>1$~MeV) of the postshock region. 
The Bernoulli parameter $b$ of the shocked fluid
then becomes substantially negative. A significant fraction of this dissociation energy
can be recovered if nucleons recombine into $\alpha$-particles \citep{bethe96}. But for this to occur, a
decrease in the temperature is required and thus the shock must
expand significantly beyond the
radius at which it typically stalls ($\sim 100-150$~km).

One of the primary goals here, and in a previous paper
\citep{FT08a} [hereafter Paper I], is to gauge the relative importance of these effects
in setting the stage for a successful explosion.  The persistence
and amplitude of a dipolar oscillation can only be reliably measured in
fully three-dimensional simulations (there are preliminary indications
that it is less prominent in three spatial dimensions than in axial symmetry; \citealt{iwakami08}).  
On the other hand,
the interplay between $\alpha$-particle recombination and hydrodynamical instabilities
has received little attention.  Although recombination is certainly present
in previous numerical studies which employ finite-temperature equations 
of state (EOSs), it should be kept in mind that considerable uncertainties in the EOS
remain at supranuclear densities.  A softening or hardening of the EOS
feeds back on the position of the shock for a given pre-collapse stellar
model \citep{marek07}.  The parametric 
study of the critical neutrino luminosity by
\citet{murphy08} is based on a single EOS;  they obtain
explosions in which the shock
seems to break out from nearly the same radial position at $\sim 250-300$~km.
Variations in the density profile of the progenitor star will similarly
modify the position of the shock, the concentration of $\alpha$-particles 
below it, and the critical neutrino luminosity for an explosion.

In this paper, we study the interplay between non-spherical shock 
oscillations, neutrino heating, and $\alpha$-particle recombination,
when the heating rate is pushed high enough for an explosion to occur.
Our focus is on the stalled shock phase, between $\sim 100$ ms and 1 s after bounce.
In a one-dimensional
calculation, the accretion flow reaches a quasi-steady state during
this interval, and the shock gradually recedes 
(e.g. \citealt{liebendoerfer01,buras06a}).

Our approach is to introduce EOSs of increasing complexity into one- and two-dimensional, 
time-dependent hydrodynamic simulations.
To this end, we use the code FLASH2.5 \citep{fryxell00},
which is well tested in problems involving
nuclear energy release in compressible fluids \citep{calder02}.
We adopt a steady state model as our initial condition, and a constant mass
accretion rate, neutrino luminosity, and fixed inner boundary.
The steady-state approximation to the stalled shock phase was
first introduced by \citet{burrows93}, and has recently been used by
\citet{ohnishi06} to study the
non-linear development of the shocked flow with a semi-realistic equation of
state and neutrino heating.
In Paper I we modeled the accretion flow as a polytropic fluid, from which
a fixed dissociation energy is removed immediately below the shock, and allowed for
neutrino cooling but not heating.  

Here we generalize this model to allow both for heating,
and for nuclear statistical equilibrium (NSE) 
between neutrons, protons, and $\alpha$-particles.
Heating is introduced in a simple, parametrized way, without
any attempt at simulating neutrino transport.  
The nuclear abundances 
are calculated as a function of pressure $p$ and density $\rho$,
using a complete finite-temperature, partially degenerate EOS.
Our model for the shocked material retains one significant 
simplification from Paper I:  we
do not allow the electron fraction $Y_e$ to vary with position below
the accretion shock.
Here there are two competing effects:
electron captures tend to reduce $Y_e$, whereas absorption of $\nu_e$ and $\bar\nu_e$
tends to drive high-entropy material below the shock toward $Y_e \simeq 0.5$.  
Since we are interested especially in the dynamics of this high-entropy material,
we set $Y_e = 0.5$ 
when evaluating the $\alpha$-particle abundance.  
To obtain a realistic density profile, we continue to approximate the internal energy
of the fluid as that of a polytropic fluid with a fixed
index $\gamma = {4\over 3}$. In reality,
the equation of state between the neutrinosphere and the shock depends in a complicated
way on the degeneracy of the electrons and the effects of electron captures.
The consequences of introducing these
additional degrees of freedom will be examined in future work.

In spite of these simplifications, our results already show many similarities 
with more elaborate collapse calculations.  
Spherical explosions are due to a global instability resembling
the one-dimensional Standing Accretion Shock Instability (SASI), 
but modified by heating. 
As in Paper I, we find that the period of the $\ell = 0$
mode remains close to twice the post-shock advection time.
Strong deformations of the shock in two-dimensional runs are
driven by material with positive Bernoulli parameter, which generally resides outside the
radius $r_\alpha$ where the gravitational binding energy of an $\alpha$-particle is equal to
its nuclear binding energy.  The recombination of $\alpha$-particles plays a major
role in creating this positive-energy material, but for this to happen the shock must
be pushed
beyond $\sim 200$ km from the neutronized core.

In this paper, we consider only
neutrino heating as the impetus for the initial expansion of the shock, rather than more
exotic effects such as rotation or magnetic fields.  
We find that the critical heating rate is a 
strong function of the initial position of the shock with respect to $r_\alpha$, 
which implies that a much higher neutrino
luminosity is needed to revive a shock that stalls well inside $r_\alpha$.  
The difference in the critical heating rate between one- and two-dimensional simulations also
depends on the size of the shock, 
and thence on the structure of the forming neutron star.

We find that the shock develops a dipolar oscillation with a large amplitude
only when the heating rate is also high enough to trigger a strong
convective instability.
We therefore surmise that buoyant motions driven by neutrino heating play a major
role in driving the dipolar oscillations that are seen in more complete simulations
of core collapse.  
Some evidence is found that acoustic wave emission by the convective
motions also can play a role.
We investigate the possibility of a heat engine within the gain
layer (the region with a net excess of neutrino heating over cooling).  
We find that most of the heat deposition by neutrinos is concentrated
in lateral flows
at the base of the prominent convective cells.  At the threshold
for an explosion, neutrino heating
plays a key role in pushing material to positive Bernoulli parameter,
but only if the shock starts well inside $r_\alpha$.

The plan of the paper is as follows.  Section \ref{s:model} presents our numerical setup, treatment of
heating, cooling, and nuclear dissociation, and outlines the sequences of models.
Sections \ref{s:1d} and \ref{s:2d} show results from 
one- and two-dimensional simulations, respectively.  We
focus on the relative effectiveness of neutrino heating and $\alpha$-particle recombination in
driving an explosion, and the relation between Bernoulli parameter and large-scale deformations 
of the shock.  The critical heating rate for an explosion is analyzed in \S\ref{s:thresholds},
and the competition between advective-acoustic feedback and convective instability is
discussed in \S\ref{s:turbulence}.  We summarize our findings in \S\ref{s:summary}.
The appendices contain details about our EOS and the numerical setup.

\section{Numerical Model}\label{s:model}

As in Paper I, the initial configuration
is a steady, spherically symmetric flow onto a gravitating point mass $M$.
The flow contains a standing shock wave, and the settling flow 
below the shock cools radiatively in a narrow
layer outside the inner boundary of the simulation volume.  
The space of such models is labeled basically by three parameters:
accretion rate $\dot M$, luminosity $L_\nu$ in electron neutrinos
and anti-neutrinos, and the radius $r_*$ of the base of the settling
flow, which corresponds roughly to the neutrinosphere radius.  
The mass $M$ of the collapsed material represents 
a fourth parameter, but it covers a narrower range than the other three.
The infalling material is significantly de-leptonized before
hitting the shock only in the first 50 ms or so of the collapse
(e.g. \citealt{liebendoerfer01}).  

We explore a two-dimensional surface through this three-dimensional
parameter space by i) fixing the ratio of $r_*$ to the initial shock radius 
$r_{s0}$ in the absence of heating ($r_*/r_{s0} = 0.4$); ii) allowing $r_{s0}$
to vary with respect to an appropriately chosen physical radius;
and then iii) increasing the level of heating until an explosion is
uncovered.  In the full problem, the shock radius at zero heating
is a unique function of $\dot M$ and $r_*$, with a small additional
dependence on $M$ and the composition of the flow outside the shock
\citep{HC92}.  The secular cooling of the collapsed core forces
a gradual decrease in $r_*$, and $\dot M$ also varies with time and with
progenitor model.  

Given the important role that $\alpha$-particle 
recombination plays in the final stages of an explosion, we implement
ii) by referencing
$r_{s0}$ to the radius where the gravitational binding energy
of an $\alpha$-particle equals its nuclear binding energy,
\begin{equation}
\label{eq:r_alpha}
r_\alpha = \frac{GMm_\alpha}{Q_\alpha} \simeq 254 M_{1.3}\textrm{ km}.
\end{equation}
Here $Q_\alpha \simeq 28.30$~MeV is the energy needed to break up 
an $\alpha$-particle into 2$n$ and 2$p$, 
$m_\alpha$ the mass of an $\alpha$-particle, and 
$M_{1.3} = M/(1.3M_\sun)$.
Choice i) allows us to consider models that have,
implicitly, both a range of physical values of $r_*$ and a range
of $\dot M$.  It is, of course, made partly for computational
simplicity (the limited size of the computational domain) and also
to facilitate a comparison between models that have different 
values of $r_{s0}/r_\alpha$.  
Nuclear dissociation is taken into account either by removing a fixed
specific energy $\varepsilon$ right below the shock,
or by enforcing NSE between $n$, $p$, and $\alpha$ throughout the 
settling flow.  Once this choice is made, the
normalization of the cooling function is adjusted to give $r_*/r_{s0} = 0.4$.
The heating rate remains freely adjustable thereafter.

We adopt this simplification because we do not intend to find the precise 
value of the critical neutrino luminosity, but instead to probe the behavior of the system around this 
critical point, whatever its absolute value.

We now describe the key components of this model in more detail,
and explain the setup of the hydrodynamic calculations.
As in Paper I, the time evolution is carried out using the second-order,
Godunov-type, adaptive-mesh-refinement code FLASH2.5 \citep{fryxell00}.

\subsection{Initial Conditions}

The introduction of heating causes a change in the structure
of the initial flow configuration.  The radius $r_s$ of the shock in the
time-independent solution to the flow equations increases with
heating rate; that is, $r_s \geq r_{s0}$.  
The material above the shock is weakly bound to the protoneutron star, and in practice
can be taken to have a zero Bernoulli parameter 
\begin{equation}\label{eq:bern}
b = \frac{1}{2}v^2 + \frac{\gamma}{\gamma-1}\frac{p}{\rho}- \frac{GM}{r}.
\end{equation}
Here $v$ is the total fluid velocity, which is radial in the initial
condition, $p$ is the pressure, $\rho$ is the mass density, and
$G$ is Newton's constant.  
The flow upstream of the shock is adiabatic and has 
a Mach number $\mathcal{M}_1=5$ at a radius $r = r_\mathrm{s0}$. 

The composition of the fluid is very different upstream
and downstream of the shock.  Changes in internal energy due to nuclear
dissociation and recombination are taken into account using the two models
described in \S\ref{s:Saha}.  
For the internal energy density of the fluid $e$, we continue
to use the polytropic relation $e = p/(\gamma-1)$; 
hence the second term on the
right-hand side of equation (\ref{eq:bern}).  Because we are not explicitly including 
changes in electron fraction due to weak processes, we keep $\gamma = 4/3$ 
for both models of nuclear dissociation.  This largely determines the density profile
inside a radius $\sim {1\over 2}r_\alpha$, where the $\alpha$-particle abundance
is very low.

The upstream and downstream flow profiles are connected through the Rankine-Hugoniot jump conditions,
which are modified so as to allow a decrement $\varepsilon$ in $b$ across the shock. 
The resulting compression factor is (Paper I)
\begin{eqnarray}
\label{eq:kappa_phot}
\kappa \equiv \frac{\rho_2}{\rho_1} & = &
(\gamma+1)\Bigg[ \left(\gamma + \mathcal{M}_1^{-2}\right) -\nonumber\\
 & &\left.\sqrt{ \left(1-\mathcal{M}_1^{-2}\right)^2 + (\gamma^2-1)\frac{2\varepsilon}{v_1^2}}
\quad \right]^{-1},
\end{eqnarray}
which reduces to $\kappa \to (\gamma+1)/(\gamma-1)$ for $\mathcal{M}_1\to \infty$ and $\varepsilon = 0$.
Throughout this paper, 
the specific nuclear dissociation energy $\varepsilon$ is defined to be
positive.  The subscripts
$1$ and $2$ denote upstream and downstream variables, respectively. 

All flow variables are made dimensionless by scaling radii to
$r_{s0}$, velocities to $v_{\rm ff\,0} = (2GM/r_{s0})^{1/2}$, 
timescales to $t_{\rm ff\,0} = r_{s0}/v_{\rm ff\,0}$, and 
densities to the upstream density at $r=\rs0$, $\rho_1(\rs0)$ [equation~\ref{eq:rho_1}].
See Paper I for further details. Throughout the paper we denote the average of a function
$F(X,...)$ over some variable $X$ by $\langle F\rangle_X$.

\subsubsection{Nuclear Dissociation}
\label{s:Saha}

We model nuclear dissociation in two ways. First, we remove a fixed specific energy
$\varepsilon$ right below the shock, as done in Paper I.   This represents the prompt and complete breakup of whatever 
heavy nuclei are present in the upstream flow.  The main limitation of this approximation is
that the dissociation energy does not change with the radius (or inclination) of the shock.
The main advantage is simplicity:  $\varepsilon$
is independent of any dimensional parameters and can be expressed as a
fraction of $\vff2$.

We also use a more accurate dissociation model which allows for NSE between $\alpha$-particles
and nucleons.\footnote{Although heavier nuclei can begin to recombine 
once the shock moves significantly beyond $r_\alpha$, this generally occurs only after 
the threshold for an explosion has been reached, and makes a modest additional
contribution to the recombination energy.}
During the stalled shock phase of core-collapse supernovae, the shock sits at $r\sim 100-200$~km, 
with a postshock temperature $T > 1$~MeV and density 
$\rho \gtrsim 10^9$~g~cm$^{-3}$. In these conditions, the heavy nuclei 
flowing through the shock are broken up into $\alpha$, $p$, and $n$.

A range of
isotopes are present in the iron core of a massive star as well as in nuclear burning shells \citep{woosley02}, but since the binding energy per
nucleon varies only by $\sim 10\%$  
we simply assume a single type of nucleus in the upstream
flow. We focus here on the later stages of the stalled shock 
phase, during which the oxygen shell is accreted.
An energy $Q_{\rm O} \simeq 14.44$~MeV must be injected to dissociate an $^{16}{\rm O}$ nucleus
into 4 $\alpha$-particles \citep{audi03}, 
which corresponds to the specific dissociation energy
\begin{equation}
\varepsilon_{\rm O} = \frac{Q_{\rm O}}{m_{\rm O}}
 \simeq 0.038 M_{1.3}^{-1} \left({r\over 150~{\rm km}}\right)\,v_{\rm ff}^2(r).
\end{equation} 
Here $m_{\rm O} \simeq 16 m_u$ is the mass of an oxygen nucleus, 
with $m_u$ the atomic mass unit.
The smallness of this number indicates that little
oxygen survives in the post-shock flow, and so we set 
the equilibrium mass fraction of oxygen to zero
below the shock, $X_{\rm O}^{\rm eq} = 0$.
The binding energy of an $\alpha$-particle is of course much larger, giving
\begin{equation}
\varepsilon_\alpha = \frac{Q_\alpha}{m_\alpha} \simeq  0.297 M_{1.3}^{-1}\left({r\over 150~{\rm km}}\right)
\,v_{\rm ff}^2(r).
\end{equation}

We find that $\alpha$-particles appear in significant numbers only
at relatively large radii ($\ga 0.5\,r_\alpha$) and in material that has 
either i) been significantly heated by
electron neutrinos closer to the neutrinosphere; or ii)
been freshly shocked outside $r_\alpha$.
The electrons are only mildly degenerate in material that has a high
entropy and $\alpha$-particle content, 
so that neutrino heating drives $Y_e$ close to $\sim 0.5$ 
(or even slightly above: see, e.g., \citealt{buras06b}).  We therefore set $Y_e = 0.5$
in the Saha equation that determines the equilibrium
mass fractions $X_n^{\rm eq}$, $X_p^{\rm eq}$ and $\Xeq = 1-X_n^{\rm eq}-
X_p^{\rm eq}$.  These
quantities are tabulated as functions of $p$ and $\rho$ using an
ideal, finite-temperature and partially
degenerate equation of state for electrons and nucleons; see 
Appendix~\ref{s:saha_detail} for details.  Specific
choices must then be made for the parameters 
$r_{s0}$, $M$, and $\dot M$; we generally take
$M = 1.3\,M_\odot$ and $\dot M = 0.3\,M_\odot$ s$^{-1}$, but allow $r_{s0}$ to vary.   An investigation
of how changes in $Y_e$ feed back onto the formation of $\alpha$-particles
is left for future work.

A specific energy
\begin{equation}
\label{eq:energy_release_general}
e_\mathrm{nuc} = - X_{\rm O} (\varepsilon_{\rm O} + \varepsilon_\alpha)
 - \left( X_\alpha - X_\alpha^\mathrm{eq}[\rho,p] \right) \varepsilon_\alpha,
\end{equation}
is either released or absorbed within a single time step (it can be of
either sign).   Here $X_{\rm O}$ is non-vanishing only for fluid elements
that have just passed across the shock, and we have set $X_{\rm O}^{\rm eq}=0$.
The quantity (\ref{eq:energy_release_general})
is introduced as an energy source term in FLASH, and
from it one readily obtains a rate of release of nuclear binding energy
per unit mass,
\begin{equation}\label{eq:energy_release_general2}
{\totd e_{\rm nuc}\over \totd t} \equiv {e_{\rm nuc}\over \Delta t},
\end{equation}
where $\Delta t$ is the simulation time step.

\begin{figure}
\includegraphics*[width=\columnwidth]{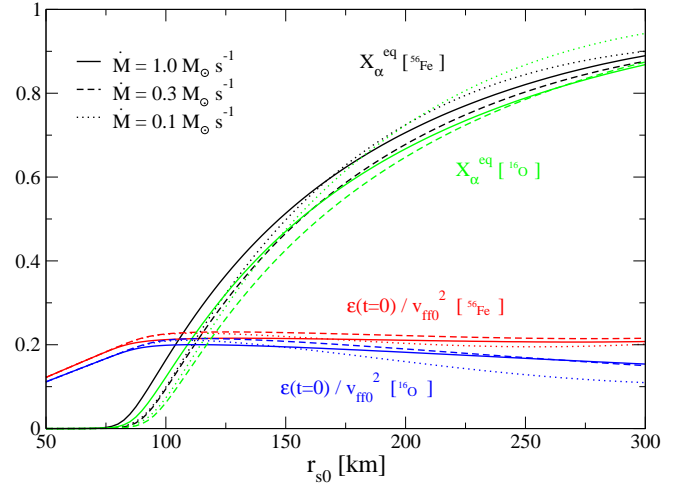}
\caption{Equilibrium mass fraction of $\alpha$-particles $\Xeq$, and ratio of initial
dissociation energy $\varepsilon(t=0)$ [equation~\ref{eq:epsilon_initial}] to $v_{\rm ff0}^2$ behind a spherical shock
positioned at radius $\rs0$.  Curves of different shadings correspond to different
mass accretion rates.
The Rankine-Hugoniot shock jump conditions and dissociation energy are calculated
self-consistently, as described in Appendix~\ref{s:ode_initial}. Square brackets refer to
the upstream composition of the accretion flow, which for simplicity is taken
to be pure ${}^{56}$Fe or ${}^{16}$O. The Mach number upstream of the shock is $\mathcal{M}_1 = 5$,
and the central mass is $M = 1.3M_\sun$.
}
\label{f:Xalpha_epsilon}
\end{figure}

In the initial condition, the dissociation energy at the shock is obtained from 
equation~(\ref{eq:energy_release_general}) using $X_{\rm O} = 1$ and $X_\alpha = 0$
upstream of the shock:
\begin{equation}
\label{eq:epsilon_initial}
\varepsilon (t=0) = \varepsilon_{\rm O} + \left(1 - X_\alpha^\mathrm{eq}[\rho_2,p_2]\right)\varepsilon_\alpha.
\end{equation}
Figure~\ref{f:Xalpha_epsilon} shows how $\varepsilon(t=0)$ and $\Xeq$ depend on the shock radius $\rs0$,
for upstream flows composed\footnote{In the case where the upstream flow is pure $^{56}$Fe, 
we replace $\varepsilon_{\rm O}$ in equation~(\ref{eq:epsilon_initial}) with 
$\varepsilon_\mathrm{Fe} = Q_\mathrm{Fe}/m_\mathrm{Fe}
\simeq 0.093 M_\mathrm{1.3}^{-1}(r/150~{\rm km})\,v_{\rm ff}^2(r)$, 
and set the electron fraction to $Y_e = 26/56$ in the 
NSE calculation behind the shock.} 
of pure $^{16}$O and $^{56}$Fe, and for different values of $\dot{M}$.
The dissociation energy is approximately constant inside
$\sim 75$ km, where the downstream flow is composed of free nucleons,
but decreases at greater distances, remaining $\sim 40\%$ of the gravitational binding energy at the shock.
The mass fraction of $\alpha$-particles reaches 50\% at $r=150-175$ km, with
a weak dependence on $\dot{M}$.

\subsubsection{Heating and Cooling in the Post-shock Flow}
\label{s:heatcool}

To allow direct comparison with our previous results,
we employ a cooling rate per unit volume of the form 
\begin{equation}
\label{eq:cooling_function}
\mathscr{L}_{C} = C p^a \rho^{b-a},
\end{equation}
with $a=1.5$, $b=2.5$, and $C$ a normalization constant.
As in Paper I,
we include a gaussian entropy cutoff to prevent runaway cooling.   
The exponents in equation (\ref{eq:cooling_function})
represent cooling dominated
by the capture of relativistic, non-degenerate electrons and positrons on
free nucleons (e.g., \citealt{bethe90}).  
Inside the radius where the
electrons become strongly degenerate, and $\alpha$-particles
are largely absent, one has $\mathscr{L}_C \propto
p_e^{3/2} n_p \propto (Y_e \rho)^3$.  
This gives essentially
the same dependence of $\mathscr{L}_C$  on $r$ as equation
(\ref{eq:cooling_function}) when $Y_e =$ constant and $\gamma = {4\over 3}$
(corresponding to $\rho \propto r^{-3}$ in a nearly adiabatic settling
flow).  In more realistic collapse calculations, $Y_e$ grows with radius
between the neutrinosphere and the shock, but
$\rho$ tends to decrease more rapidly than $\sim r^{-3}$
(e.g. \citealt{buras06a}).  
Our chosen form for the cooling function results in a slightly
wider gain region and, therefore, a slightly lower critical
heating rate for an explosion.
The bulk of the cooling occurs in a narrow layer close to the accretor at $r=r_*$, and
the accreted material accumulates in the first few computational
cells adjacent to the inner boundary without
a major effect on the rest of the flow.

We model neutrino heating as a local energy generation rate per unit volume of the form
\begin{equation} 
\label{eq:heating_function}
\mathscr{L}_H = H (1-X_\alpha) \rho /r^2.
\end{equation} 
The normalization constant $H$ measures 
the strength of the heating.  
The factor $(1-X_\alpha)$ accounts for the fact that the cross section for neutrino 
absorption by
$\alpha$-particles is much smaller than that for free nucleons \citep{bethe90}. 
For simplicity, we do not include the flux factor due to the transition between diffusion and 
free-streaming.
Our focus here is on the nature of the instabilities occurring in the
flow near the threshold for an explosion, and we do not attempt a
numerical evaluation of the critical heating rate.

An additional energy source term arises from the change in the equilibrium fraction of
$\alpha$-particles as they are advected in the steady state initial solution. The instantaneous 
adjustment of $X_\alpha$ to its equilibrium value, combined with 
equation~(\ref{eq:energy_release_general}), 
yields an energy generation rate per unit volume
\begin{equation}
\label{eq:alpha_advection}
\mathscr{L}_\alpha  =  \rho v \varepsilon_\alpha \frac{\totd X^\mathrm{eq}_\alpha}{\totd r}\nonumber \\
                    =  \rho v \varepsilon_\alpha \left[ 
                         \frac{\partial \Xeq}{\partial \rho}\frac{\totd \rho}{\totd r} + 
                         \frac{\partial \Xeq}{\partial p}\frac{\totd p}{\totd r}\right].
\end{equation}
This energy generation rate is negative, as the temperature increases inwards and thus the $\alpha$-particle fraction decreases with decreasing radius ($v$ is negative).

\subsubsection{Numerical Setup}
\label{s:setup}

In our time dependent calculations, we use one-dimensional and two-dimensional spherical coordinates with baseline
resolution $\Delta r_\mathrm{base} = \rs0/320$ and $\Delta \theta_\mathrm{base} = \pi/192$, with one 
extra level of mesh refinement inside $r = r_* + 0.1(\rs0-r_*)$ to better resolve the steep density 
gradient that arises in the cooling layer. We do not employ a hybrid Riemann solver because we do not 
see the appearance of the odd-even decoupling instability \citep{quirk94}.

We employ a reflecting inner boundary condition at $r=r_*$ for the sake of simplicity; we do not attempt 
to model the protoneutron star (as done by \citealt{murphy08}) or its contraction through a moving inner 
boundary [as done by \citet{scheck06} and \citet{scheck08}]. The outer boundary condition is kept fixed
at $r = 7\rs0$, and is set by the upstream flow at that position.

To trigger convection below the shock, we introduce random cell-to-cell velocity perturbations in $v_r$ and $v_\theta$
at $t=0$, with an amplitude $1\%$ of the steady state radial velocity. To study
the interplay between shock oscillations and convection, we also drop overdense shells with a given
Legendre index $\ell$, as done in Paper I, without random velocity perturbations.

In order to track the residency time of the fluid in the gain region, we assign a scalar
to each spherically symmetric mass shell in the upstream flow. This scalar is passively advected by 
FLASH2.5. Through this technique,
we are able to assign a ``fluid" time to each element in the domain, corresponding to the time at which the
mass shell would cross the instantaneous angle averaged shock position if advected from the outer boundary at
the upstream velocity:
\begin{equation}
\label{eq:t_f}
t_F = t_{\rm OB} + \int_{\langle r_s(t)\rangle_\theta}^{r_{\rm OB}} \frac{\totd r}{|v_r|}.
\end{equation}
Here $t_{\rm OB}$ is the time at which the fluid enters through the outer radial boundary at $r=r_{\rm OB}$, and
$\langle r_s(t)\rangle_\theta$ is the angle averaged shock position. Initially, $t_{\rm OB} = 0$ and all the 
fluid below the shock is set to $t_F = 0$. This prescription works well for statistical studies (\S\ref{s:residency}),
tracing large scale fluid patches, despite some inevitable turbulent mixing on small scales.

An explosion is defined as either i) a collision between the shock and 
the outer boundary of the simulation volume ($r=7\rs0$)
within $1000t_{\rm ff0}$ of the start of the simulation;
or ii) in the special case of the one-dimensional constant-$\varepsilon$
models, a transient expansion that breaks a quasi-steady pattern 
within the same timeframe.
Even in the spherically symmetric simulations, very small changes in heating rate can lead
to dramatic changes in shock behavior, and so this definition of explosion is good enough for our purposes.

\begin{figure}
\includegraphics*[width=\columnwidth]{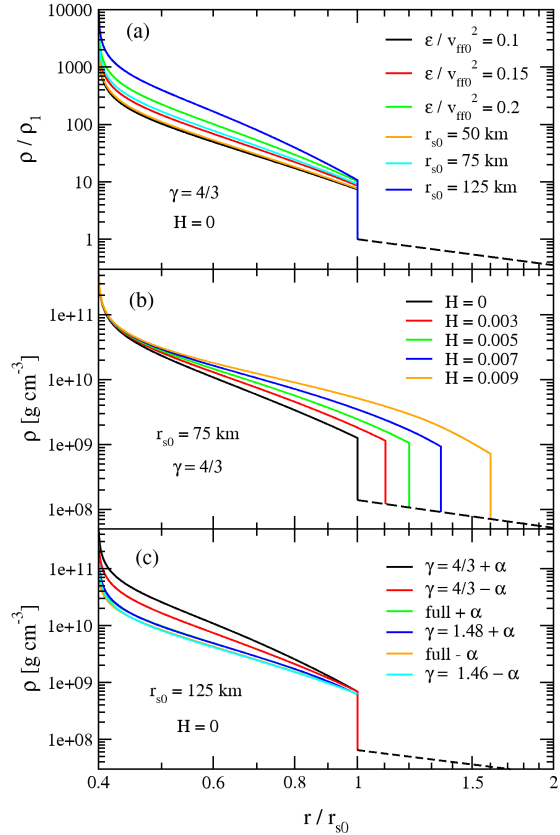}
\caption{Sample initial density profiles, which are solutions to
the spherically symmetric and time-independent flow equations.
Panel (a) shows the zero-heating configurations for all the sequences
listed in Table~\ref{t:setups}.
Other parameters are $\{\gamma=4/3$, $\mathcal{M}_1(\rs0) = 5\}$ for all
configurations, and  $\{\dot{M} = 0.3M_\sun$~s${}^{-1}$, $M = 1.3M_\sun\}$ for the
NSE models.
Panel (b) shows a sequence
with a fixed cooling function and range of heating rates
($H$ is given in units of $\rs0 v_{\rm ff\,0}^3$). The dashed line
shows the upstream flow.
Panel (c) shows a sequence with different equations of
state, $r_{s0} = 125$~km, and $H=0$.
The labels ``$+\alpha$'' and ``$-\alpha$'' mean with and
without $\alpha$-particles included in the EOS, while ``full'' means
that the EOS explicitly includes finite-temperature and partially degenerate
electrons, black body photons, and ideal-gas ions.
All other parameters are the same as in (a). Only the $\gamma=4/3$ upstream
flow is shown.
See Appendix~\ref{s:ode_initial} for further details.
}
\label{f:profiles}
\end{figure}

\subsection{Model Sequences}\label{s:model_seq}

We choose six sequences of models, each with a range of heating parameters $H\geq 0$, and
each evolved both in spherical and axial symmetry. Their parameters are summarized in 
Table~\ref{t:setups}.   In each sequence, the normalization of the cooling 
function is chosen so that $r_*/r_s = 0.4$ 
at zero heating.  Three sequences have a constant
dissociation energy, which take the values $\varepsilon/\vff2=\{0.1,0.15,0.2\}$.  
The other
three sequences assume NSE below the shock, and have shock radii $\rs0=\{50,75,125\}$~km at zero heating.
This means that the physical value of the cooling radius also takes on different values, namely
$\{20,30,50\}$~km.  In effect, our models are probing different sizes for the neutrinosphere,
and different times following the collapse.
The other parameters in the NSE models are $M=1.3M_\sun$ and $\dot{M}=0.3M_\sun$~s$^{-1}$.

Table~\ref{t:setups} samples some properties of a few models from each sequence:  one 
with zero heating, another with $H$ close to the critical value for an explosion, and a third
with the largest heating parameter that will allow a steady solution.
Note that the shock starts out at $\sim 1.3\,\rs0$ in the time-independent,
spherical flow solution, and quickly saturates at $\sim (1.8-2)\rs0$ in the two-dimensional models
with heating just below threshold for an explosion.
The quantity $\varepsilon/v_1^2$ references the dissociation energy to (twice)
the kinetic energy of the upstream flow, and is the key free parameter
determining the compression rate $\kappa$ across the shock (equation [\ref{eq:kappa_phot}]). 

When examining how  the prescription for nuclear dissociation influences
the results, we will focus on
the $\varepsilon=0.15\vff2$ sequence and the NSE sequence with
$\rs0=75$~km, which have similar initial density profiles (due 
to the low initial $\alpha$-particle abundance in the NSE model).

The six initial models at zero heating are shown in  
Figure~\ref{f:profiles}a. Panel (b) shows the sequence of initial models
with $r_{s0} = 75$km and a range of heating parameters.  The model with
$H=0.007v_{\rm ff0}^3\rs0$ is close to the threshold for an explosion,
while the one with
$H=0.009v_{\rm ff0}^3\rs0$ is well above threshold.  At higher values of $H$, cooling by 
$\alpha$-particle dissociation (equation~[\ref{eq:alpha_advection}]) 
can be significant in a layer below the shock, causing the 
density profile to steepen slightly.

Fig~\ref{f:profiles}c shows how our constant-$\gamma$,
ideal gas approximation to the internal energy of the flow
compares with the full EOS containing finite-temperature and 
partially degenerate
electrons (see Appendix \ref{s:ode_initial} for details).  
The curves labeled ``$+\alpha$'' include our prescription
for heating/cooling by $\alpha$-particle recombination/dissociation,
and those labeled ``$-\alpha$'' do not.  We show the sequence
with the largest shock radius ($r_{s0} = 125$ km) so that NSE allows some
$\alpha$'s to be present.
The neglect of electron captures below the shock results in
an adiabatic index between $4/3$ and $5/3$ in the zone where $\alpha$-particles
are absent.  This causes the EOS to stiffen, so that the density profile
is well approximated by an ideal gas with $\gamma\simeq 1.48$ at zero heating.
Adding in heating tends to flatten the density profile even more, and
with $\gamma = 1.48$ it would be much flatter than is typically seen in
a realistic core collapse model.   Hence we choose an EOS with $\gamma = 4/3$.

\begin{deluxetable}{ccccccc}
\tablecaption{Sample Configurations\label{t:setups}}
\tablewidth{0pt}
\tablehead{
\colhead{$\varepsilon/\vff2$} &
\colhead{$H v_{\rm ff0}^{-3}\rs0^{-1}$} &
\colhead{$r_{\rm s}/\rs0$} &
\colhead{$\varepsilon/v_1^2$} &
\colhead{$\kappa$} &
\colhead{$\chi$} &
\colhead{$\phantom{\Xeq}$}
}
\startdata
0.1  & 0                         & 1.00 & 0.10 & 7.3  & 0   & \\
     & 8.00E-3                   & 1.27 & 0.13 & 7.7  & 4.5 & \\
     & 1.48E-2\tablenotemark{*}  & 2.57 & 0.31 & 11.0 & 22  & \\
0.15 & 0                         & 1.00 & 0.15 & 8.6  & 0   & \\
     & 7.00E-3                   & 1.29 & 0.20 & 9.6  & 9.0 & \\
     & 1.17E-2\tablenotemark{*}  & 2.34 & 0.41 & 18.9 & 40  & \\
0.2  & 0                         & 1.00 & 0.20 & 10.1 & 0   & \\
     & 5.50E-3                   & 1.30 & 0.27 & 12.5 & 19  & \\
     & 8.38E-3\tablenotemark{*}  & 2.08 & 0.47 & 34.9 & 74 & \\ 
\noalign{\smallskip}
\hline
\hline
\noalign{\smallskip}
\colhead{$\rs0$~[km]} & 
\colhead{$H v_{\rm ff0}^{-3}\rs0^{-1}$} & 
\colhead{$r_{\rm s}/\rs0$} &
\colhead{$\varepsilon(t=0)/v_1^2$} & 
\colhead{$\kappa$} & 
\colhead{$\chi$} &
\colhead{$X_\alpha^{\rm eq}(r_s)$}\\
\noalign{\smallskip}
\hline
\noalign{\smallskip}
50  & 0                         & 1.00 & 0.11 & 7.6  & 0   & 5.5E-6 \\
    & 8.00E-3                   & 1.29 & 0.15 & 8.1  & 5.5 & 6.1E-5 \\
    & 1.43E-2\tablenotemark{*}  & 3.01 & 0.26 & 8.6  & 27  & 0.43   \\ 
75  & 0                         & 1.00 & 0.17 & 9.0  & 0   & 4.3E-4 \\
    & 6.50E-3                   & 1.30 & 0.22 & 10.1 & 11  & 4.5E-2 \\
    & 1.15E-2\tablenotemark{*}  & 3.61 & 0.21 & 6.8  & 51  & 0.83   \\ 
125 & 0                         & 1.00 & 0.21 & 10.7 & 0   & 0.26   \\
    & 3.50E-3                   & 1.33 & 0.21 & 9.9  & 22  & 0.51   \\
    & 7.28E-3\tablenotemark{*}  & 3.99 & 0.18 & 6.1  & 120 & 0.99   \\
\enddata
\tablenotetext{*}{Maximum heating rate for a steady flow solution \citep{burrows93}.}
\end{deluxetable}

\section{One Dimensional Simulations}\label{s:1d}

\subsection{Shock Oscillations and Transition to Explosion}
\label{s:1d1}

An explosion in spherical symmetry involves the development of
an unstable $\ell = 0$ SASI mode. We showed in Paper I that,
in the absence of neutrino heating, the period of
this mode is essentially twice the post-shock advection time.  
As heating is introduced into the flow, we find that this
relation is maintained.  The $\ell = 0$ mode is damped until the
heating rate is pushed above a critical value, which we now 
discuss.
\begin{figure}
\includegraphics*[width=\columnwidth]{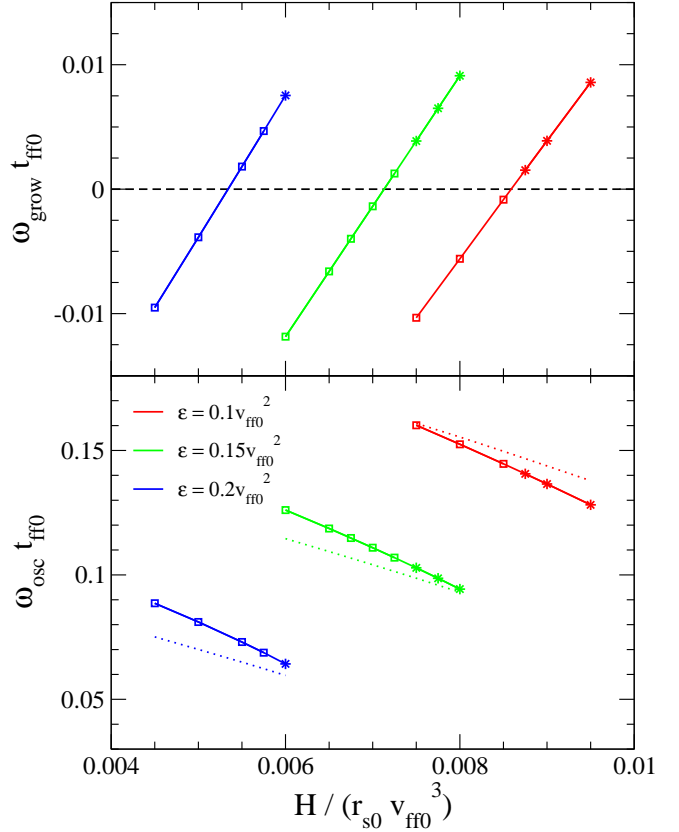}
\caption{Linear growth rates (top) and oscillation frequencies (bottom) of 
one-dimensional models with constant dissociation energy $\varepsilon$,
as a function of heating parameter $H$ around the threshold for explosion.
Stars denote configurations
that explode within $1000t_{\rm ff0}$. Increased heating makes the system
more unstable because the density profile flattens, akin to an increase in $\gamma$.
Dotted lines show the frequency $\omega_{\rm osc} = 2\pi/(2 t_{\rm adv})$. 
Oscillation frequencies decrease with increasing heating rate because $r_s$ moves
out relative to $r_*$, so that the advection time $t_{\rm adv}$ (equation [\ref{eq:tadv}])
increases.
Increasing the dissociation energy raises the oscillation period,
and so a somewhat higher heating rate is required to obtain an
explosion in a finite interval.
}
\label{f:linear_growth_1d}
\end{figure}

It should be emphasized that this critical heating rate
is generally lower than that defined by \citet{burrows93},
which marked the disappearance of a steady, spherically symmetric
solution to the flow equations.
Large amplitude shock oscillations in spherical symmetry have been witnessed near the threshold for 
explosion in calculations by \citet{ohnishi06} and \citet{murphy08}.  Both calculations
employed a realistic EOS, but like us
included neutrino heating as a local source 
term in the energy equation. Oscillations have also been seen by \citet{buras06b} in more
elaborate calculations with Boltzmann neutrino transport.  

The origin of the spherically symmetric SASI oscillation can be briefly summarized as follows.
An initial outward shock displacement generates 
an entropy perturbation, which is negative for $\gamma \la 5/3$.
This entropy perturbation is advected down to the cooling layer,
where it causes, at constant ambient pressure, an increase
in the cooling rate,
$\delta \mathscr{L}_C/\mathscr{L}_C = -[(\gamma-1)/\gamma](b-a) \delta S > 0$.
The resulting negative pressure perturbation is rapidly communicated to the shock,
which recedes and generates an entropy perturbation of the opposing sign.
One more iteration results 
in a shock displacement of the same sign as the initial
displacement, and allows the cycle to close.  
The duration of the $\ell = 0$ mode is therefore nearly twice the advection 
time from the shock to the cooling layer,
\begin{equation}\label{eq:tadv}
{2\pi\over \omega_{\rm osc}} \simeq 2\int_{r_*}^{r_s} {\totd r\over |v_r|}.
\end{equation}
The cycle is stable
for $\gamma=4/3$ and $r_*/\rs0=0.4$.
\begin{figure}
\includegraphics*[width=\columnwidth]{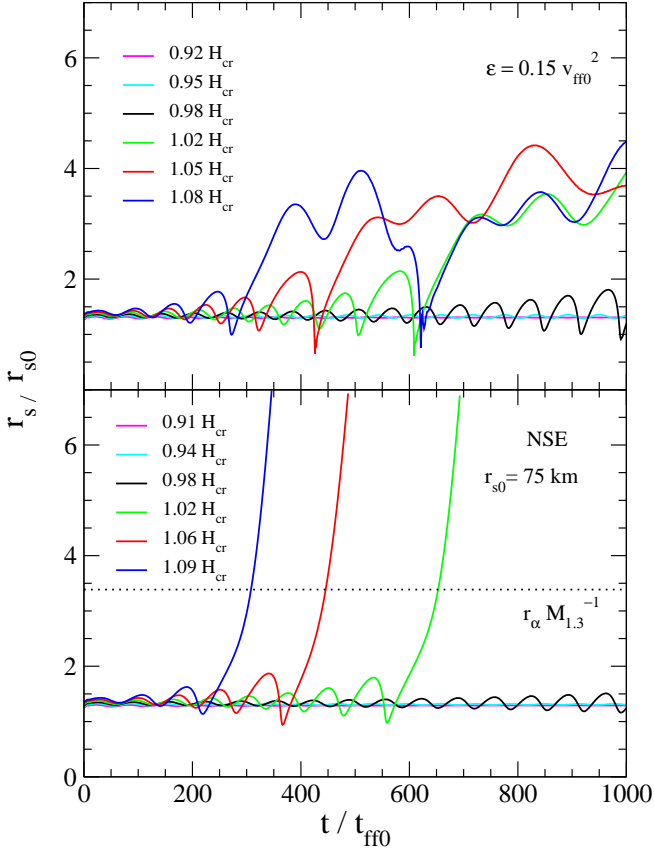}
\caption{Shock radius as a function of time for two sequences of one-dimensional simulations. 
The upper panel shows runs with constant dissociation energy $\varepsilon=0.15\vff2$,
and a range of heating coefficients $H$ near the critical value $H_{\rm cr} = (0.006625\pm 0.000125)v_{\rm ff\,0}^3\rs0$.
The lower panel shows runs with $\rs0 = 75$~km
that include recombination of $\alpha$-particles. 
In this case, $\Xeq$ is initially negligible everywhere below the shock (see Table~\ref{t:setups}), but
grows as the shock expands.  The horizontal dotted line labels the radius
$r_\alpha$ at which the nuclear binding energy $Q_\alpha$ of an $\alpha$-particle
equals its gravitational binding energy (equation~\ref{eq:r_alpha}).
The critical heating for this second sequence is lower,
$H_{\rm cr} = (0.006125\pm 0.000125)v_{\rm ff\,0}^3\rs0$.
}
\label{f:rshock_threshold_1d}
\end{figure}

When heating is added, the density profile flattens. 
Increasing  $ \gamma $  has the same effect, and has been found to push up the growth 
rate of linear SASI modes (e.g. Paper I).
There is a critical heating rate for which the damping effect of the spherically symmetric SASI is neutralized and 
there is no net growth. We find that, once the heating rate exceeds this critical value, 
the system always explodes. 

We therefore define the critical heating rate in our spherically symmetric simulations 
to be the minimum heating rate for growing shock oscillations.\footnote{We define
our critical heating parameter $H_{\rm cr}$ to be the average
of the values in the exploding and non-exploding runs that are 
closest to the threshold for explosion, within our fiducial $1000t_{\rm ff0}$ cutoff.}
Figure~\ref{f:linear_growth_1d} shows real and imaginary eigenfrequencies as a function
of heating rate for our one-dimensional initial configurations with constant $\varepsilon$. 
The curves were
obtained by solving the differential system of \citet{F07}, modified to account
for a constant rate of nuclear dissociation (Paper I) as well as incorporating
the heating function in equation~(\ref{eq:heating_function}). 
The runs marked by stars explode within a time $1000 t_{\rm ff0}$, and so
require a small, but finite, positive growth rate.  

In an exploding run, the expansions become longer and contractions shorter
as the shock oscillation develops a large amplitude.
Eventually the accretion flow is halted during a contraction.
This marks the point of explosion, beyond which the feedback between
the shock and the cooling layer is broken.  
Material then tends to pile up in the gain region, is further heated, and more
material reverses direction. The net effect is to push the shock outward.
Movie 1 in the online material illustrates this chain of events.

\begin{figure}
\includegraphics*[width=\columnwidth]{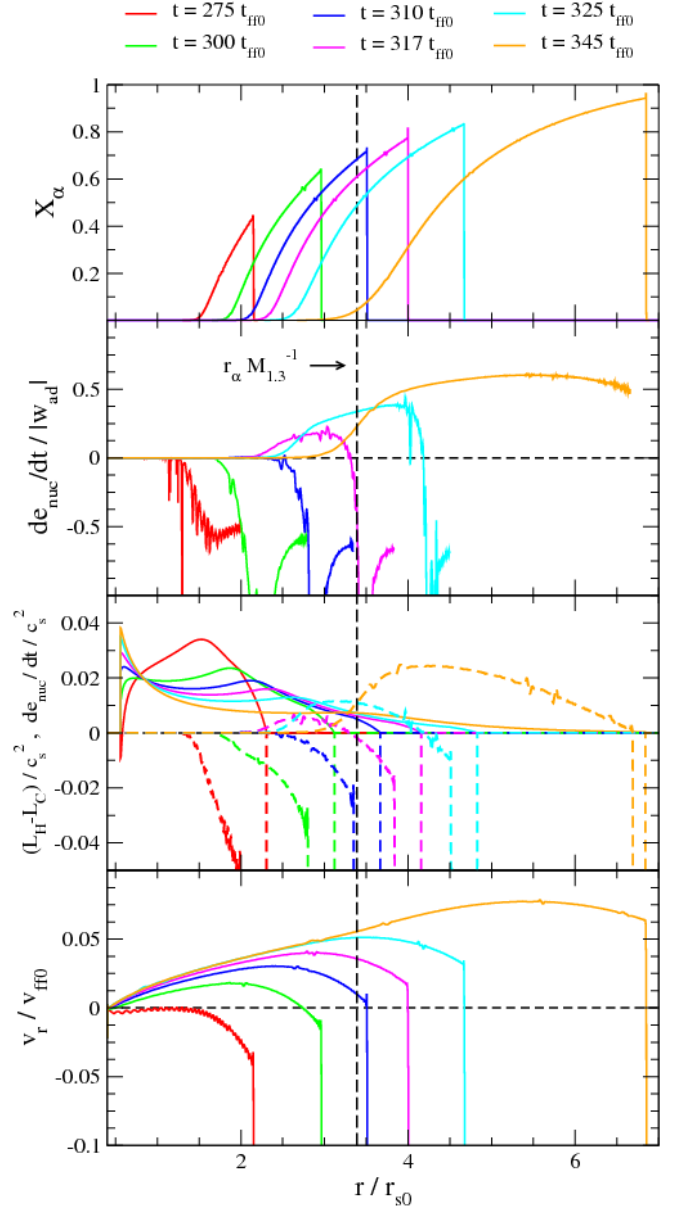}
\caption{Radial profiles of various quantities during
shock breakout in the NSE model
with $H=1.09H_{\rm cr}$ and $r_{s0} = 75$ km 
(Figure~\ref{f:rshock_threshold_1d}).
Top panel: mass fraction of $\alpha$-particles.  Second panel:
rate of release of specific nuclear binding energy $\totd e_{\rm nuc}/\totd t$
compared with the (adiabatic) rate of 
change of enthalpy
$w_{\rm ad}$ [equation~\ref{eq:wad}].  Third panel: net neutrino heating rate
per unit volume ${\mathscr L}_H-{\mathscr L}_C$ (thin
solid curves) and $\totd e_{\rm nuc}/\totd t$ (thick dashed
curves), both
normalized to the local value of $c_s^2 = \gamma p/\rho$.
Bottom panel:  radial velocity normalized to $v_{\rm ff0}$ at
radius $r_{s0}$.  
Both $\totd e_{\rm nuc}/\totd t$ and $w_{\rm ad}$ are smoothed 
in radius for clarity. 
}
\label{f:Xalpha_explosion}
\end{figure}

\begin{figure*}
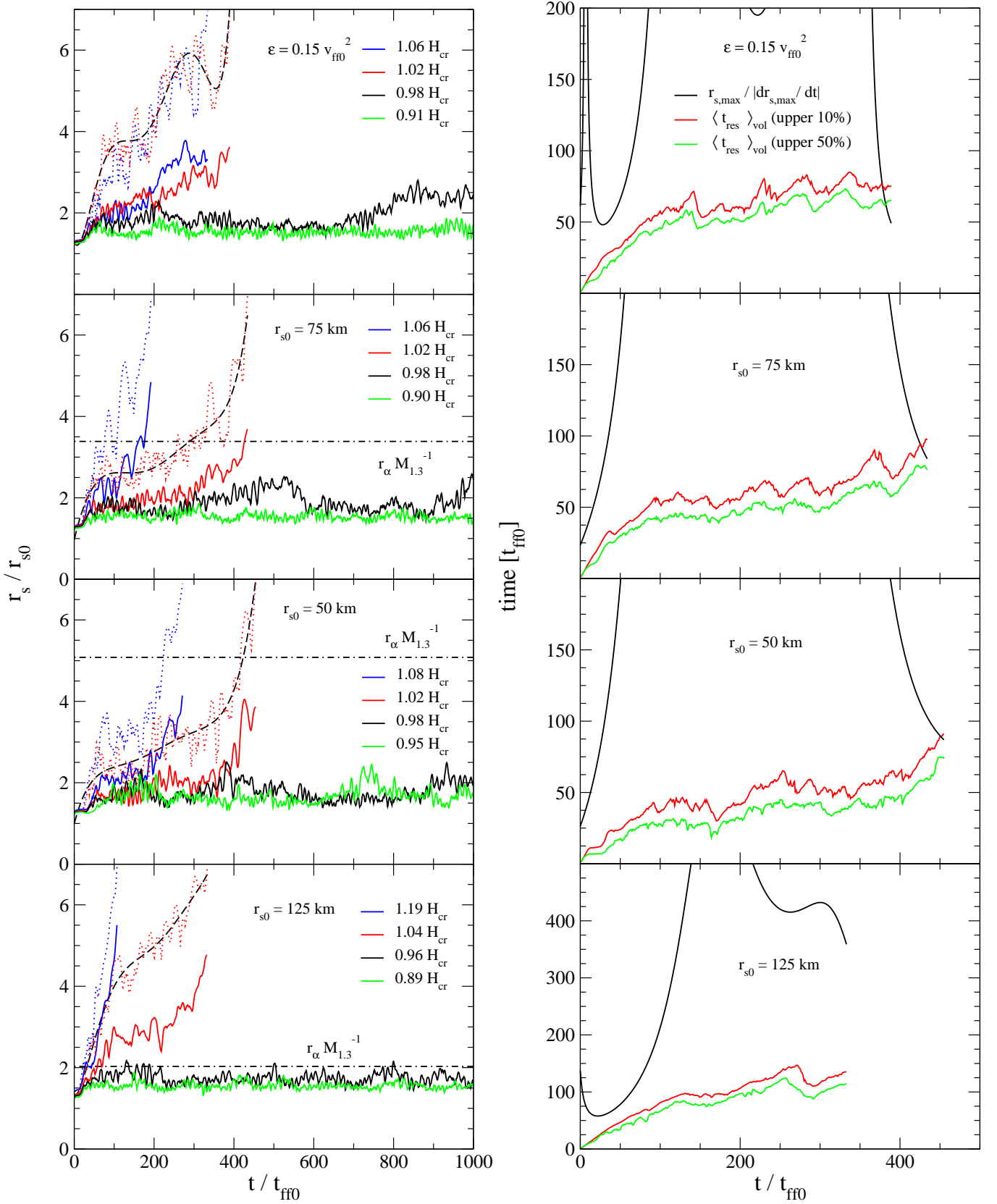

\includegraphics*[width=0.5\textwidth]{f6a.eps}
\includegraphics*[width=0.493\textwidth]{f6b.eps}
\caption{Left panel:  Angle-averaged shock radius (solid lines) and maximum shock radius
(dotted lines) for various two-dimensional models around the threshold for explosion.
Upper panel shows runs with constant dissociation energy $\varepsilon=0.15\vff2$, while
lower panels displays NSE runs with $\rs0$ as labeled. Critical heating rates $H_{\rm cr}$
are different for each configuration, and can be found in Figure~\ref{f:thresholds_B_epsilon}.
Right panel: Black lines show expansion timescale of maximum shock 
radius $t_{\rm exp}\sim r_{\rm s,max}/|\totd r_{\rm s,max}/\totd t|$, computed using a polynomial fit
for the runs just above the threshold for explosion (corresponding to the black dashed lines
on the left panels).  
Green and red lines show the average residency time over the 50\% and 10\% of the gain
region volume with highest $t_{\rm res}$, respectively 
(see \S\ref{s:residency} for the definition of this timescale). 
Shock breakout occurs whenever $t_{\rm exp} \sim  \langle t_{\rm res}\rangle_{\rm vol}$, 
except in the model where recombination heating is
dominant ($r_{s0} = 125$ km). 
}
\label{f:rshock_threshold_2D}
\end{figure*}

\subsection{Effects of Alpha-Particle Recombination}
\label{s:alpha_1d}

Shock breakout is controlled by the build-up of positive energy fluid downstream of the shock,
and therefore is sensitive to the density profile below the shock.  Heating by neutrinos is concentrated
fairly close to the protoneutron star, inside a distance $\sim (2-3)r_*$.  
Heating by $\alpha$-particle recombination is concentrated
at a greater distance $\sim r_\alpha$ (equation [\ref{eq:r_alpha}]), 
but still can reach a comparable amplitude.  

The dependence of shock breakout on heating rate is displayed in Figure~\ref{f:rshock_threshold_1d} 
for two accretion models and several values of $H$ close to $H_{\rm cr}$ (see Table~\ref{t:setups}).
The initial expansion of the shock during the explosion phase is very similar for 
models with constant $\varepsilon$ and with NSE in the shocked fluid.
However, the time evolution bifurcates near the radius $r_\alpha$. 

Figure~\ref{f:Xalpha_explosion} shows 
successive profiles of the shocked flow in the exploding run with
$H=1.09H_{\rm cr}$ and $r_{s0} = 75$ km.  
The $\alpha$-particle fraction approaches unity as the shock reaches the radius $r_\alpha$.
The second panel shows the specific nuclear energy
generation rate [equation~(\ref{eq:energy_release_general2})] normalized to the adiabatic rate of 
change of the enthalpy,
\begin{equation}
\label{eq:wad}
w_{\rm ad} = \frac{1}{\rho}\frac{\totd p}{\totd t} = -c_s^2 \nabla \cdot \mathbf{v}.
\end{equation}
Here $c_s = (\gamma p /\rho)^{1/2}$ is the sound speed.
The third panel compares the amplitude and distribution of neutrino and recombination heating,
and the bottom panel plots the radial velocity
in the postshock region.  

We can summarize this behavior as follows:
during the initial expansion phase, fluid below the shock continues to move inward, and
the dissociation of $\alpha$-particles removes energy from the flow  (as expected from 
equation~[\ref{eq:alpha_advection}]).  
Some fluid behind the shock begins to move outward around $300\tff$,
but nuclear dissociation still causes a net loss of internal energy.
However, the recombination of $\alpha$-particles sets in
above $r_\alpha$, especially in regions where $X_\alpha \lesssim 0.5$.
By the time the shock hits the outer boundary, $\totd e_{\rm nuc}/\totd t$ exceeds 
one-half of $|w_{\rm ad}|$.

\begin{figure*}
\includegraphics*[width=0.5\textwidth]{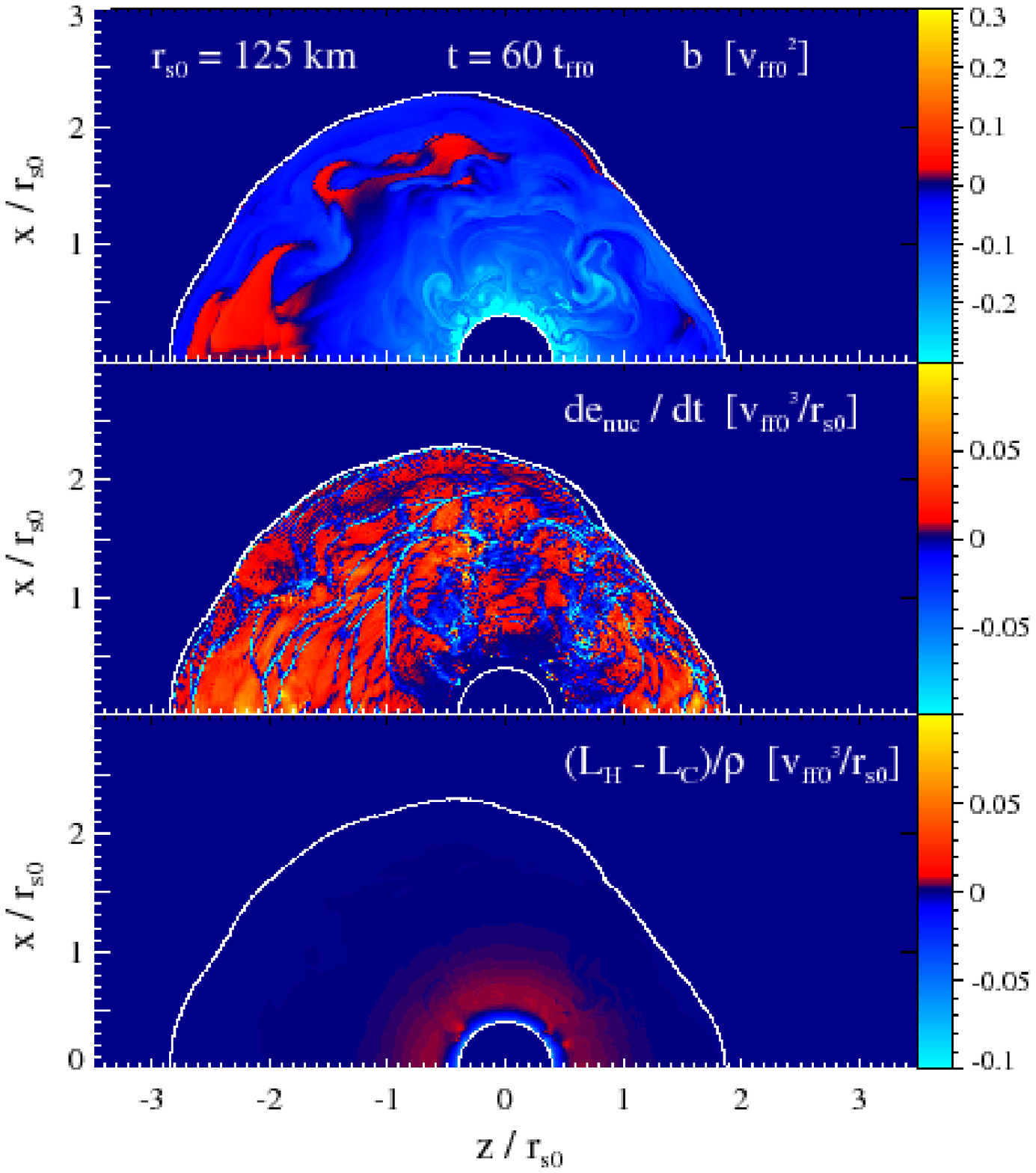}
\includegraphics*[width=0.5\textwidth]{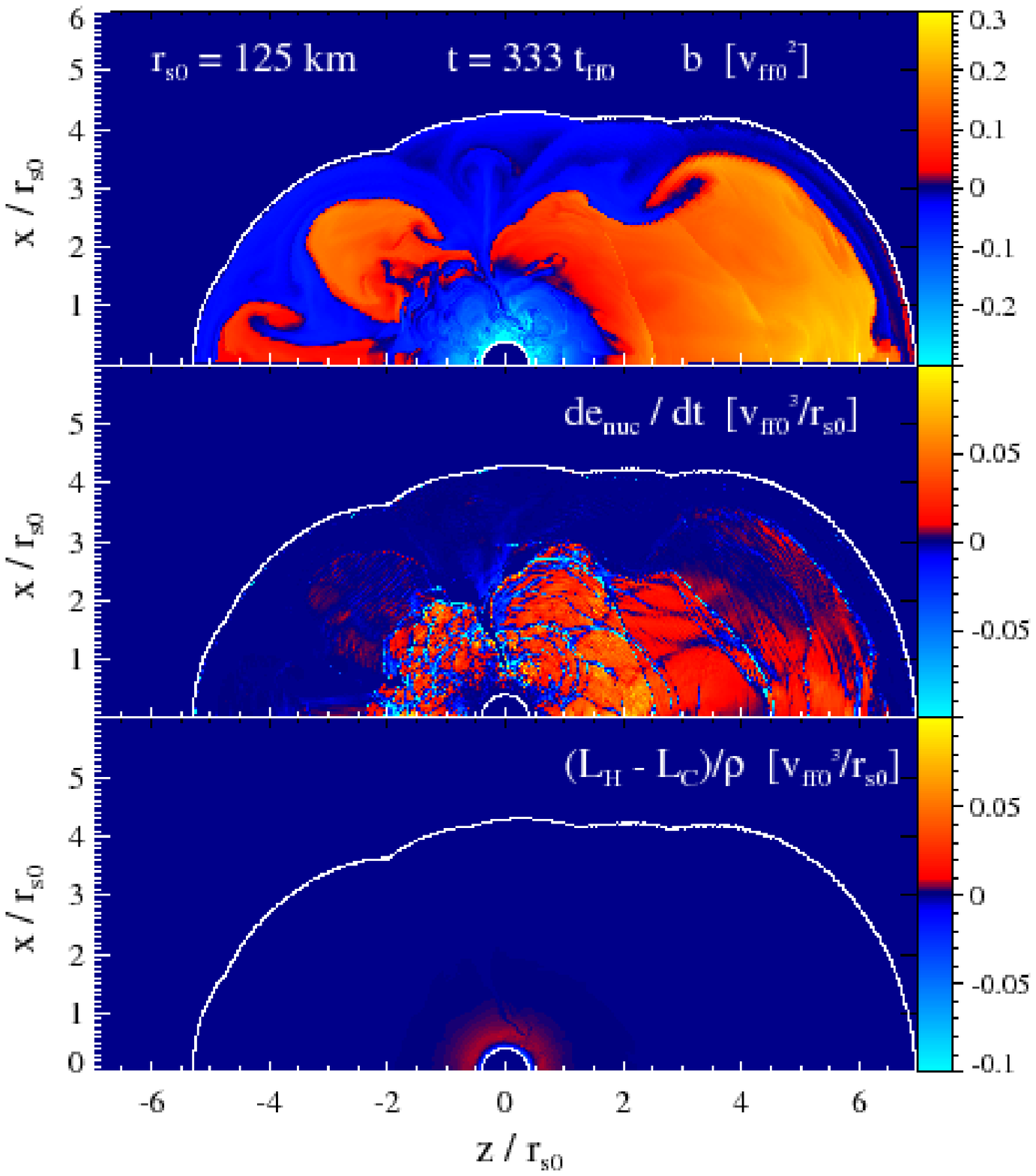}
\includegraphics*[width=0.5\textwidth]{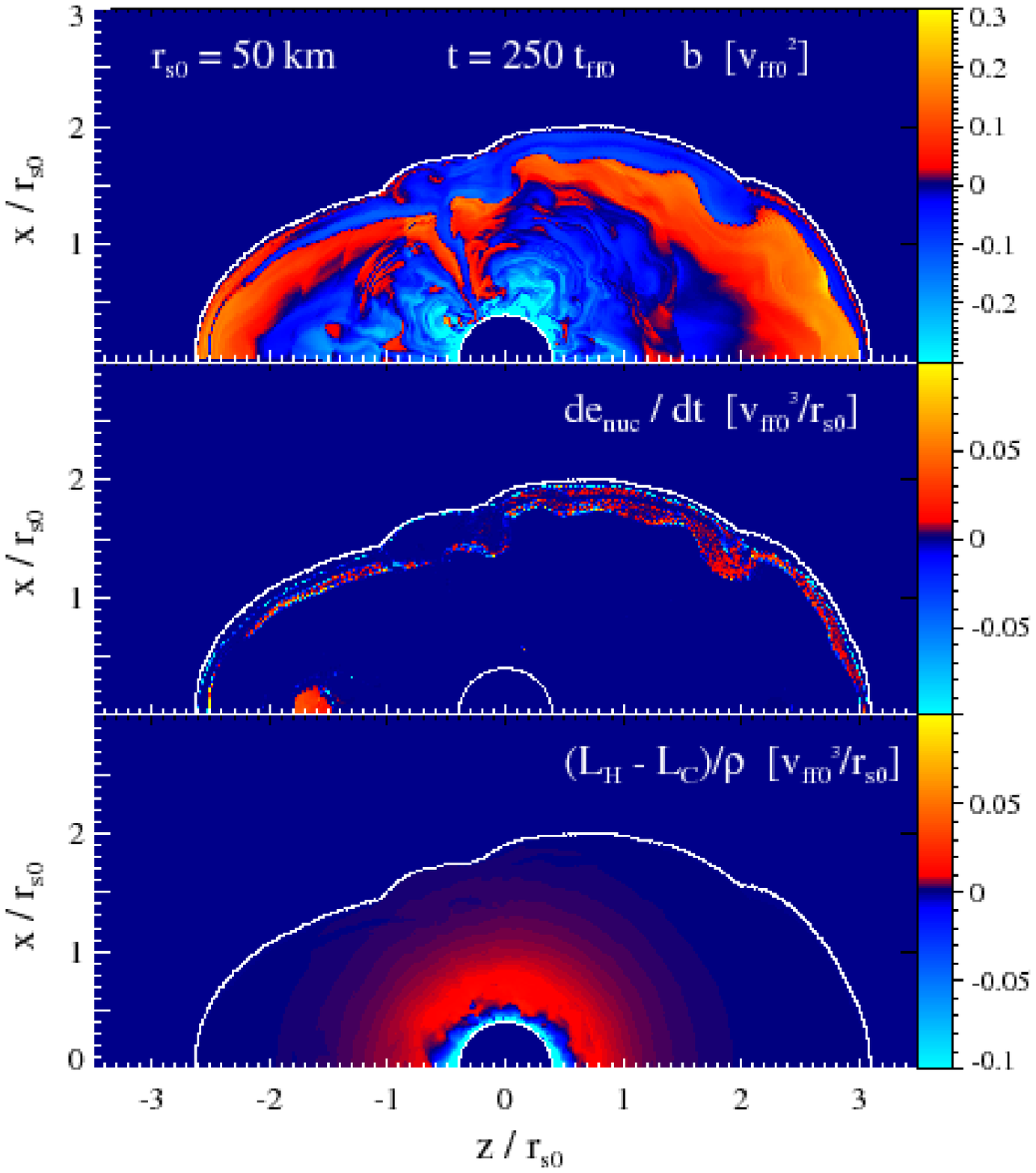}
\includegraphics*[width=0.5\textwidth]{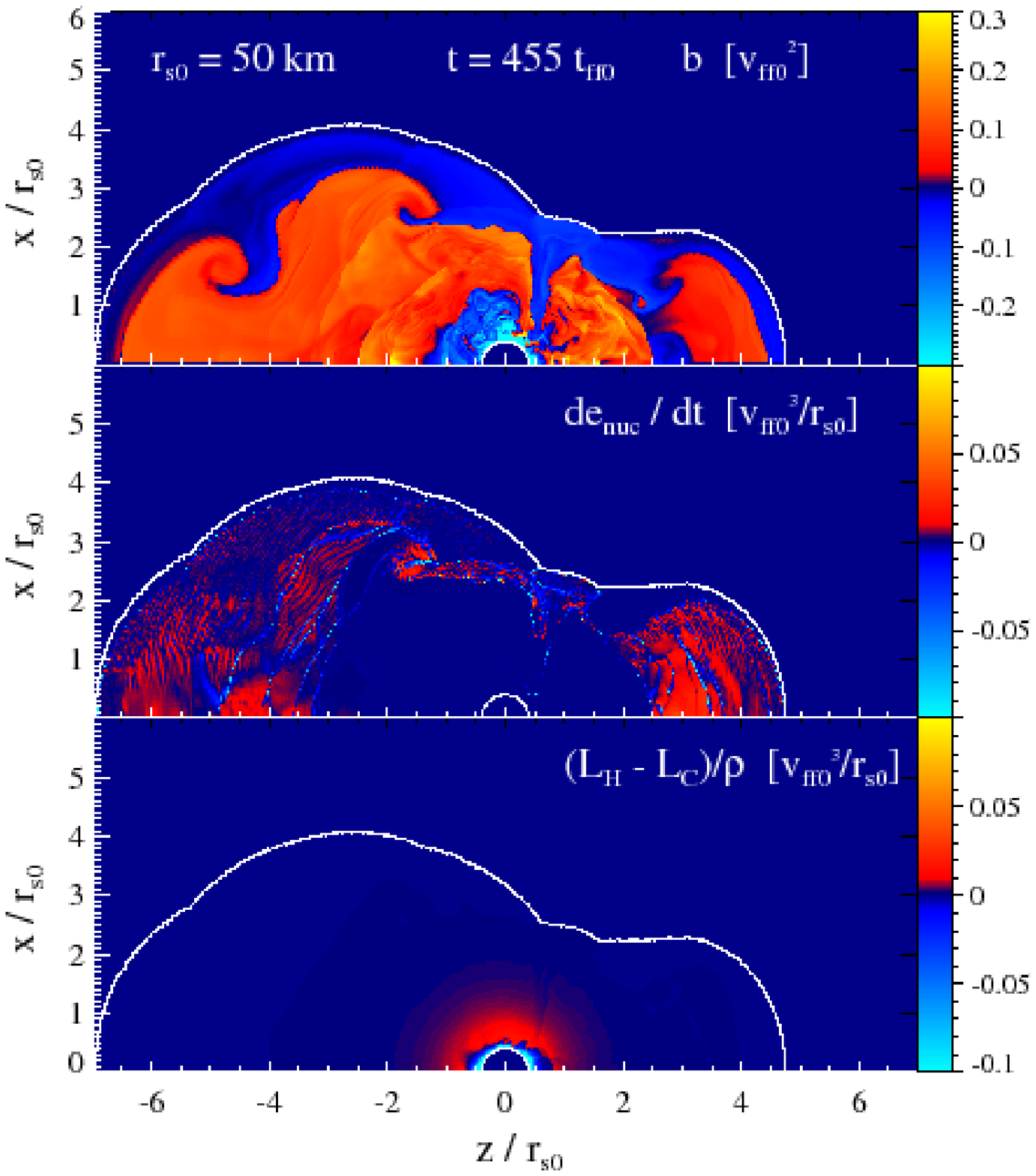}
\caption{Snapshots of two separate models, with heating parameter
$H$ just above the threshold for an explosion, and initial shock
radius either well inside $r_\alpha$ ($r_{s0} = 50$ km, without heating)
or close to $r_\alpha$ ($r_{s0} = 125$ km, without heating).
Within each panel, the top figure displays 
Bernoulli parameter $b$; the middle figure the
rate of change of nuclear energy generation;
and the bottom figure the net rate of neutrino heating.
Top left: the early development of an 
asymmetric plume with positive $b$; top right: the same run
just before the shock hits the outer boundary. 
In this $r_{s0} = 125$ km run, the 
heating by $\alpha$-particle recombination is enhanced
with respect to neutrino heating due to the large 
$X_\alpha$ in the initial stationary model. 
The central zone with $b<0$ maintains a nearly spherical
boundary near the radius $r_\alpha$ ($\simeq 2.0\,r_{s0}$),
and recombination heating straddles this boundary.   
Bottom left: $\alpha$-particles begin to form as the shock
approaches $r_\alpha$ in the $r_{s0} = 50$ km run, but neutrino
heating remains much stronger than recombination heating.
Bottom right: the same run just before the shock hits the outer boundary.
When the shock starts off well inside $r_\alpha$, neutrino heating
dominates the initial expansion, and material with $b>0$ forms
well inside $r_\alpha$ (see \S\ref{s:dynamics_2d}). 
Animations showing the evolution of these two configurations
are available in the online version of the article.
}
\label{f:energies_r125}
\end{figure*}

The dependence of the density contrast
$\kappa$ (equation [\ref{eq:kappa_phot}]) on radius also has an influence on the details of breakout.
When the dissociation energy $\varepsilon$ is held fixed, $\kappa$ increases toward larger radius.
This has the effect of creating a dense layer of fluid below the shock when 
the shock has reached a radius where
$\varepsilon \sim v_{\rm ff}^2/2$.  In spherical symmetry, the breakout of the shock is 
then impeded by this layer, which cannot exchange position with
the lighter material below it.  It can happen that
the energy in the expanding region is no longer able to sustain the heavier material above,
and the shock collapses, as shown in Figure~\ref{f:rshock_threshold_1d} for the constant-$\varepsilon$ run 
with $H = 1.08H_{\rm cr}$. This obstruction is avoided when statistical equilibrium between $n$, $p$, and $\alpha$
is maintained below the shock, because
$\varepsilon/v_1^2$ 
and $\kappa$ both decrease gradually 
as the shock expands to distances much larger than $\rs0$
(Figure \ref{f:Xalpha_epsilon}).
This limit to the shock expansion does not occur in two dimensions, as 
the superposition of dense fluid over lighter fluid is
Rayleigh-Taylor unstable on the dynamical time $t_{ff0}$.

\section{Two-Dimensional Simulations}
\label{s:2d}

Extending the flow calculation to two dimensions reveals some subtle patterns of behavior.  
The time evolution of the shock is shown in the left panel of Figure~\ref{f:rshock_threshold_2D} for a range of
heating rates near the threshold for explosion.   In contrast with the one-dimensional runs, the
breakout of the shock looks similar in models with constant dissociation energy
and with NSE between $n$, $p$, and $\alpha$ below the shock.  Both types of models are
subject to buoyancy-driven instabilities, which allow cold material
below the shock to interchange position with hotter material
within the gain region.  As a result, the shock is highly 
asymmetric at breakout in both cases.   
In \S \ref{s:thresholds} we compare the critical heating rate for
explosion in one- and two-dimensional runs, and examine how it is influenced by $\alpha$-particle recombination.  

Around the threshold for explosion, all of our runs develop vigorous convective motions
before the SASI has a chance to undergo even a few oscillations.   
At high heating rates, we find that the convective instability is 
driven by the negative entropy gradient 
within the layer of maximal neutrino heating.
In non-exploding runs, the shock settles to a quasi-equilibrium state
with oscillations taking place over a range of 
angular (Legendre) index $\ell$, as previously seen by
\citet{ohnishi06}, \citet{scheck08}, and \citet{murphy08}.  
The amplitude of the
$\ell = 1$ and 2 modes remains small until the heating parameter $H$ has begun to exceed
about one half the critical value for an explosion. 
The competition between SASI growth and convective instability is 
examined in detail in \S\ref{s:turbulence}.  

\begin{figure}
\includegraphics*[width=\columnwidth]{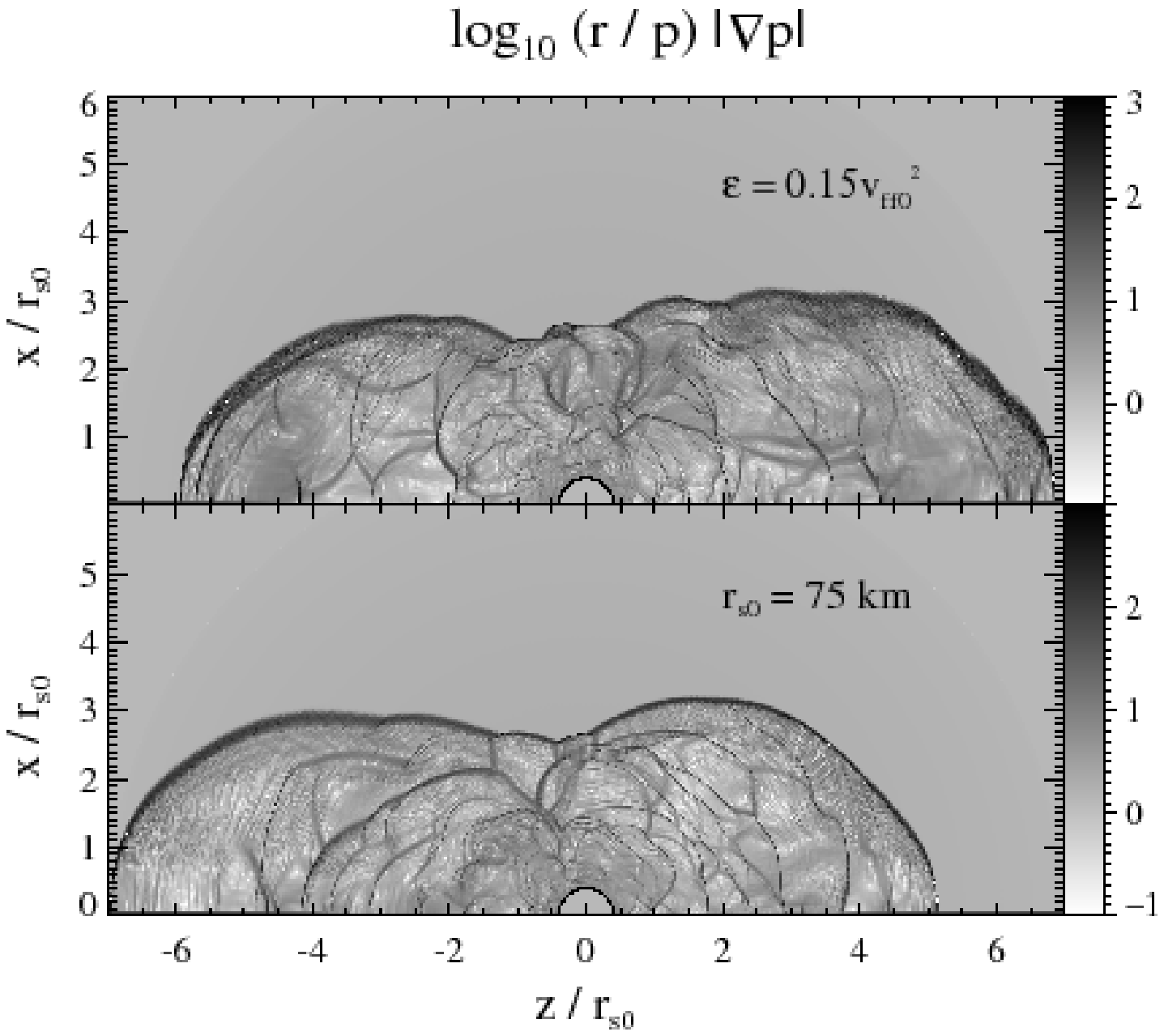}
\caption{Normalized pressure gradient $(r/p)|\nabla p|$ showing the secondary shock structure
during breakout.  The top model is $\varepsilon=0.15\vff2$ and the bottom
NSE with $\rs0=75$~km.  Both have heating rates just above the threshold for an explosion. 
An animation showing the time evolution is available in the online version of the article.
}
\label{f:secondary_shocks}
\end{figure}

\begin{figure}
\includegraphics*[width=\columnwidth]{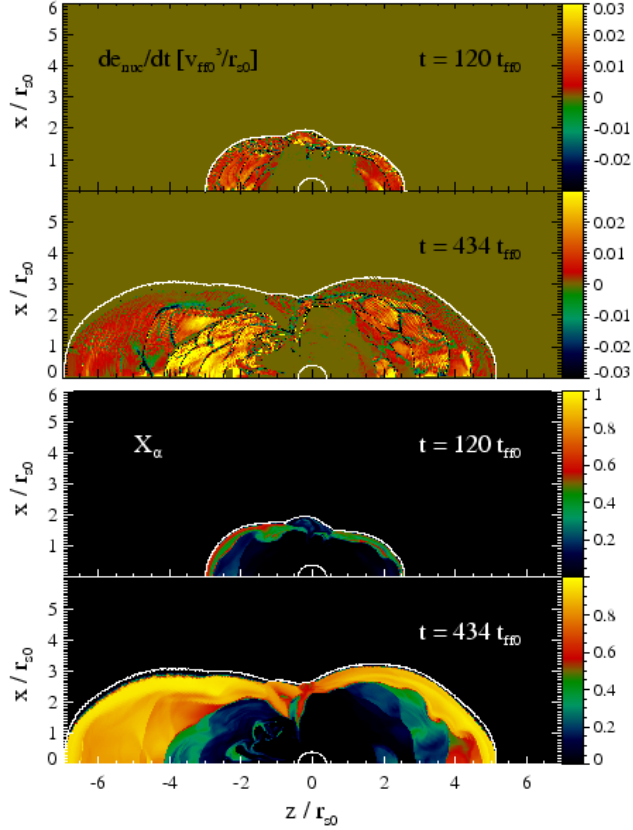}
\caption{Top panels: rate of release of specific nuclear binding energy
$\totd e_{\rm nuc}/\totd t$.  Bottom panels: mass fraction of $\alpha$-particles
$X_\alpha$.
We show two instants in the exploding NSE 
run with $\rs0 = 75$~km and $H=1.02H_{\rm cr}$. The shock contour
is approximated by the white line which marks $X_{\rm O} = 90\%$.
}
\label{f:enuc_xfue_explode}
\end{figure}

\begin{figure*}
\includegraphics*[width=\textwidth]{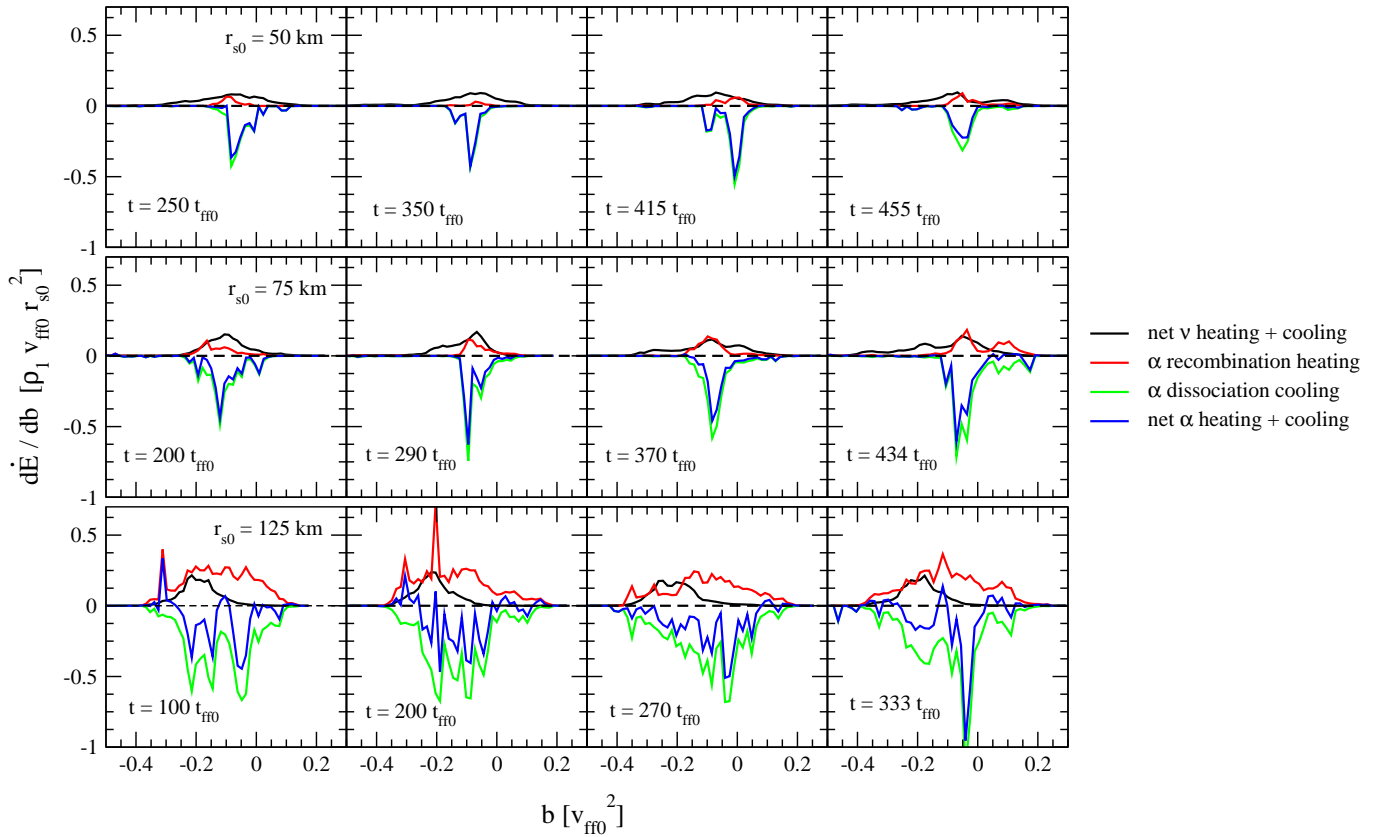}
\caption{
Heating rate of material, as distributed
with respect to Bernoulli parameter $b$.
This illustrates the relative importance of neutrino heating and nuclear 
dissociation/recombination in hot and cold parts of the flow.
We restrict attention to material in the gain region 
(defined by $\mathscr{L}_H > \mathscr{L}_C$)
in the three two-dimensional NSE runs just above the threshold for explosion. 
Four snapshots are shown: the pre-explosion quasi-steady state (leftmost),
onset of explosion (second from left to right), and breakout (third and fourth).
Black curves: net heating
rate resulting from neutrino absorption and emission.  
Red/green curves:  heating/cooling rate by $\alpha$-particle recombination
and dissociation in material with $\totd e_{\rm nuc}/\totd t > 0$ and 
$\totd e_{\rm nuc}/\totd t < 0$, respectively.
Blue curves:  net heating/cooling rate due to changing $\alpha$-particle
abundance.  The sharp negative spike near $b=0$ represents $\alpha$-particle
dissociation in fresh, cold downflows.  The formation of
material with $b > 0$ is primarily due to $\alpha$-particle
recombination in the $r_{s0} = 125$ km run.  As the initial
radius of the shock is reduced with respect to $r_\alpha$, neutrino
heating makes a proportionately larger contribution near breakout.
}
\label{f:heating_hist}
\end{figure*}

\begin{figure}
\includegraphics*[width=\columnwidth]{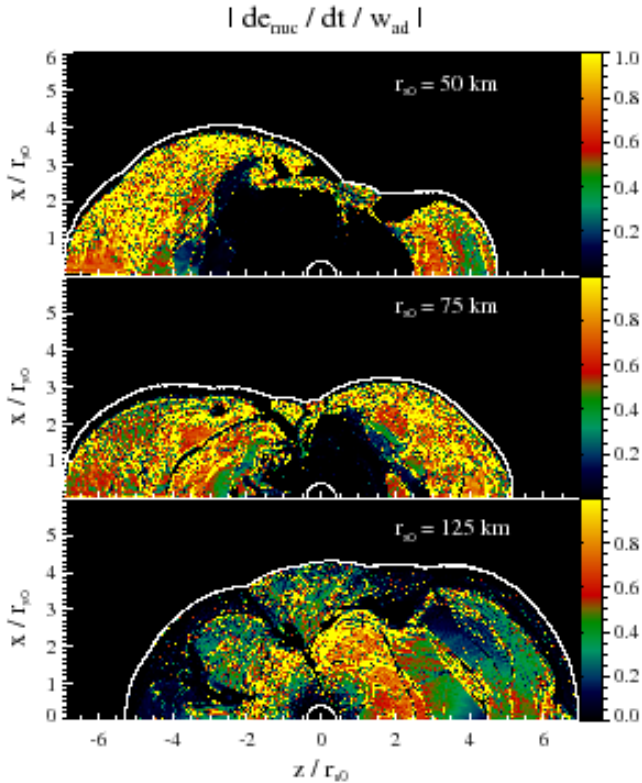}
\caption{Ratio of $\totd e_{\rm nuc}/\totd t$ (rate of release of nuclear binding energy,
equation~[\ref{eq:energy_release_general2}]) to $w_{\rm ad}$ (adiabatic rate of
change of the enthalpy,
equation~[\ref{eq:wad}]). NSE models shown have $H$ just above $H_{\rm cr}$.
}
\label{f:ephot_wad}
\end{figure}

\begin{figure*}
\includegraphics*[width=\columnwidth]{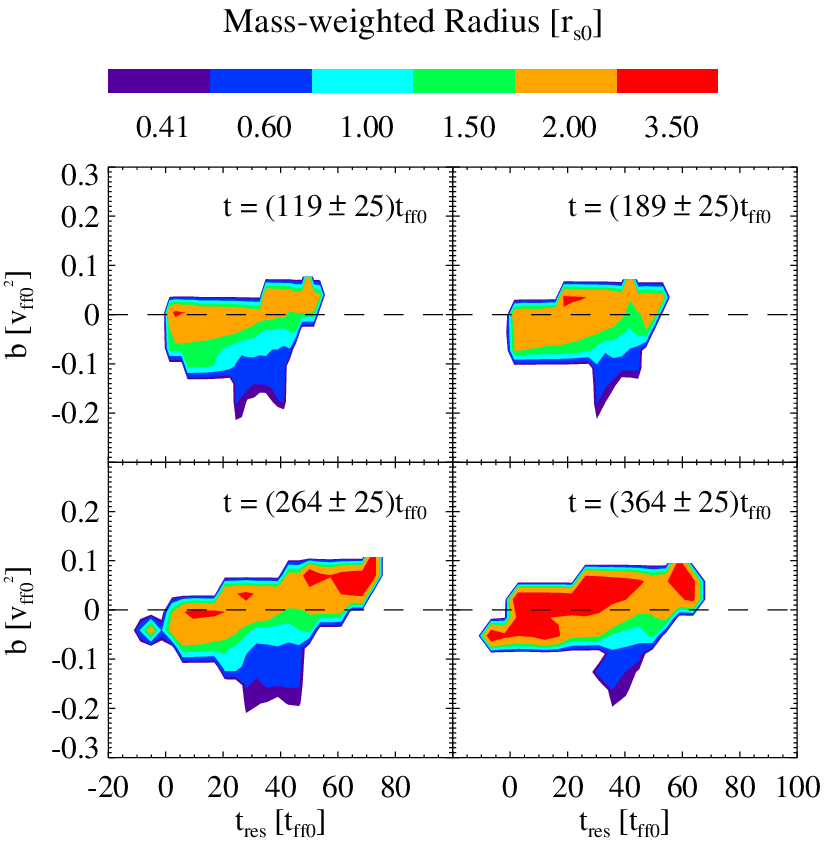}
\includegraphics*[width=\columnwidth]{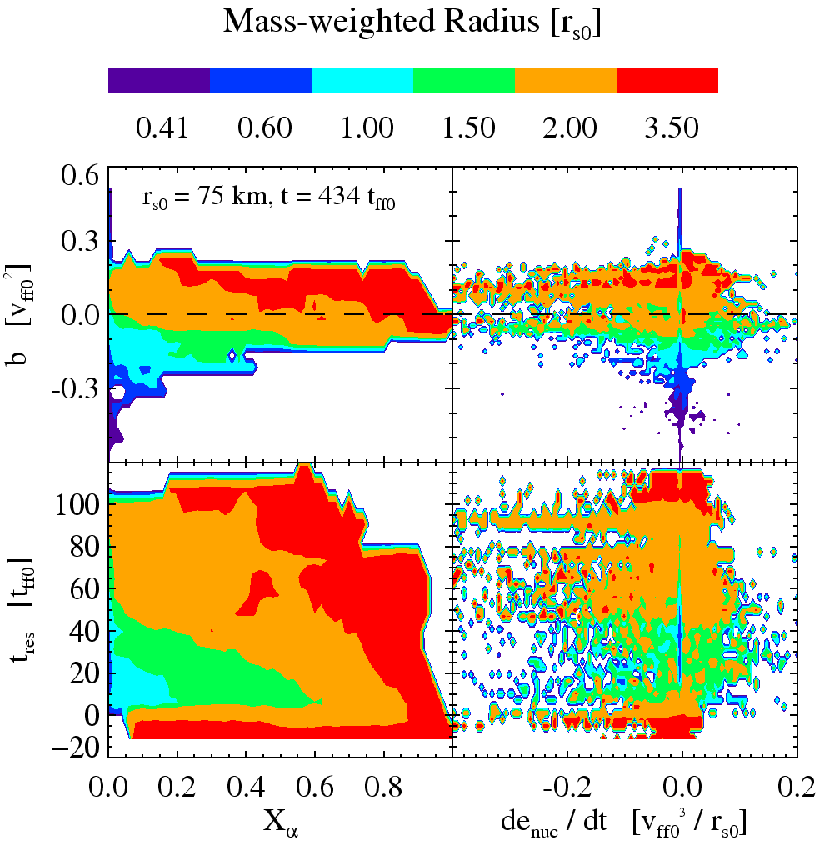}
\caption{Left panel:  Histogram of Bernoulli parameter $b$ and residency 
time $t_{\rm res}$ in the 
exploding run with $\varepsilon = 0.15\vff2$ 
(see also Figure~\ref{f:berc_explode}).  The 
colors label the mass-weighted radius, and 
we include all material experiencing
a net excess of neutrino heating over cooling.
Right panel:  Histogram of Bernoulli parameter $b$ and residency time
$t_{\rm res}$ versus $\alpha$-particle mass fraction $X_\alpha$ and 
rate of release of specific nuclear binding energy $\totd e_{\rm nuc}/\totd t$,
in the exploding run with $r_{s0} =$ 75 km
(Figure~\ref{f:enuc_xfue_explode}).
}
\label{f:histogram}
\end{figure*}

One gains considerable insight into the mechanism driving shock breakout
by examining the distribution of Bernoulli parameter (equation [\ref{eq:bern}])
in the shocked fluid.  We first consider the NSE runs with
$r_{s0} = 50$ km and $125$ km, with the heating parameter $H$
just above the threshold for an 
explosion.  Two snapshots from each of these runs are shown
in Figure~\ref{f:energies_r125}.
In the first case, the initial equilibrium shock radius
is only $\sim r_\alpha/4$ km, and $\alpha$-particles are essentially absent
below the shock.  In the second, the shock starts at $\sim 2r_\alpha/3$
and $X_\alpha \sim 0.5$ initially
in the postshock flow.  

Large deformations of the outer shock are caused by convective
plumes that carry positive energy.  Strong neutrino heating is generally
concentrated
within an inner zone where the material is gravitationally bound (b < 0). 
The degree of symmetry of this bound
material depends on the $\alpha$-particle abundance.  
In the $r_{s0} = 125$ km run, it is spherically stratified
and the material with $b > 0$ is generally excluded from it.
Strong recombination heating is present both below and
above the surface where $b\simeq 0$, indicating that
it is mainly responsible for imparting positive energy to 
the shocked material.  
The mean shock radius expands by a factor $\sim 2.5$ between the
upper two frames in Figure~\ref{f:energies_r125}, but the growth in the
volume of positive-energy material is not accompanied by a significant
expansion of the inner bound region, whose outer radius remains fixed
at $r \simeq r_\alpha$.  

This segregation of bound from unbound material is broken
when the shock is more compact initially, as is seen in the lower
two panels of Figure~\ref{f:energies_r125}.   A single dominant 
accretion plume is continuously present, which funnels cold and dense
material into the zone of strong neutrino heating.  Alpha-particles
are present only well outside the boundary between $b<0$ and $b>0$.
The competition between
$\alpha$-particle and neutrino heating is discussed in more detail
in \S\ref{s:alpha_effects},
and the influence of $\alpha$-particles on the threshold heating rate
for an explosion is examined in \S \ref{s:thresholds}.

The accumulation of a bubble of hot material right behind the shock is a consequence of the balance of the buoyancy
force acting within the bubble, and the ram pressure of the preshock material. The ratio of force densities is \citep{thompson00}
\begin{equation}
\frac{F_{\rm buoy}}{F_{\rm ram}} \simeq \left(\frac{\rho - \rho_{\rm bubble}}{\rho}\right)\left(\frac{2GM/r_s}{v_r^2}\right)
\Delta \Omega_{\rm bubble},
\end{equation}
where $r_s$ is the shock radius, $v_r$ is the ambient radial flow speed, $\rho$ is the ambient density,
$\rho_{\rm bubble}$ the density of the bubble, and $\Delta \Omega_{\rm bubble}$ is its 
angular size.  A low-density bubble ($\rho-\rho_{\rm bubble} \sim \rho$) can 
resist being entrained by the convective flow
once it grows to a size $\Delta \Omega_{\rm bubble}\sim {\cal M}_{\rm con}^2$~Sr, where
${\cal M}_{\rm con}$ is the convective Mach number. On the other hand, the 
bubble must attain a much larger angular size
$\Delta \Omega_{\rm bubble}\sim1$~Sr if the buoyancy force is to overcome the upstream
ram ($|v_r| \sim v_{\rm ff}$) and force a significant expansion of the shock surface.  
Figure~\ref{f:energies_r125} shows that
the extent of the shock expansion is indeed correlated with the angular width of the region where hot material accumulates.

Another interesting feature of Figure~\ref{f:energies_r125} is the presence of secondary shocks,
which are triggered once the outer shock becomes significantly non-spherical.
Their locations are marked by discrete jumps in the rate of recombination heating.
Secondary shocks are also prevalent throughout the nonlinear phase
in the constant-$\varepsilon$ models. 
Figure~\ref{f:secondary_shocks} shows the 
normalized pressure gradient $(r/p)|\nabla p|$ for collapse models of both types,
when $H$ is just above threshold for an explosion (right before the shock hits the 
outer boundary of the simulation volume).  The
online version of the article contains an animated version of Figure~\ref{f:secondary_shocks} showing
the complete evolution. In both cases, secondary shocks extend over the whole postshock domain, signaling
the dissipation of supersonic turbulence which is stirred by accretion plumes that penetrate
into the gain region.

\subsection{Distribution of Alpha Particle Recombination Heating}
\label{s:alpha_effects}

Heat input by neutrino absorption and by $\alpha$-particle recombination have 
very different distributions within the shocked fluid:  strong neutrino
heating is concentrated inside
$r_{s0}$, whereas recombination heating of
a comparable amplitude is distributed throughout the settling flow.
Strong recombination heating quite naturally extends below
the zone where $\alpha$-particles are present in significant numbers,
as is seen in Figure~\ref{f:enuc_xfue_explode}.  
The first and third panels of this
figure depict the pre-explosion steady state of the $\rs0 = 75$~km model with
$H = 1.02H_{\rm cr}$, 
while the second and fourth panels show the last time before the shock hits the outer boundary.
At the latter time, 
one sees that the strongest
recombination heating is concentrated in a layer where 
$X_\alpha \la 0.5$,
at the base of the extended $\alpha$-rich plumes.
Just as in the one-dimensional simulations (e.g. Figure~\ref{f:Xalpha_explosion}), 
$X_\alpha$ approaches unity during shock breakout.

The relative strength of neutrino heating and recombination heating
depends on the initial radius of the shock, and on the Bernoulli parameter
of the 
postshock material.
Figure~\ref{f:heating_hist}
separates out cooling by $\alpha$-particle dissociation from
heating by recombination and neutrino radiation
during the pre-explosion quasi-steady state (leftmost panels),
at the onset of explosion (second panel left to right),
and during breakout (two rightmost panels). See 
Figure~\ref{f:rshock_threshold_2D} for comparison.  The colored
curves show the positive, negative, and net contributions 
from nuclear energy generation.
The sharp negative spike near $b=0$ represents $\alpha$-particle
dissociation in fresh, cold downflows.  The formation of
material with $b > 0$ is primarily due to $\alpha$-particle
recombination in the $r_{s0} = 125$ km run.  As the initial
radius of the shock is reduced with respect to $r_\alpha$, neutrino
heating makes a proportionately larger contribution near breakout.

The strength of the boost given to the shock by recombination
heating can be gauged by comparing
$\totd e_{\rm nuc}/\totd t$ to the adiabatic rate of change $w_{\rm ad}$ of the 
enthalpy of the flow (equation [\ref{eq:wad}]).
Figure~\ref{f:ephot_wad} shows the result for all 
three NSE sequences with $H$ just above $H_{\rm cr}$. 
In all cases, $\totd e_{\rm nuc}/\totd t \simeq w_{\rm ad}$ in various parts of the shocked fluid 
once the shock extends beyond a radius $\simeq r_\alpha$.
Most of the heat input by recombination is concentrated where $X_\alpha \sim 0-0.5$,
just as in spherical symmetry. 

Histograms of $\totd e_{\rm nuc}/\totd t$ versus $b$ and $X_\alpha$ are shown in 
the right panel of
Figure~\ref{f:histogram}.
The rapid dissociation of $\alpha$-particles in fresh downflows is represented by
the long tail toward large negative values of $\totd e_{\rm nuc}/\totd t$,
showing that the overall contribution of nuclear energy generation is negative. 
The $\alpha$-particle concentration 
is very stratified, with higher $X_\alpha$ occurring at larger
radius. 
Most of the mass with positive Bernoulli parameter is located at large radii. 

\begin{figure*}
\includegraphics*[width=\textwidth]{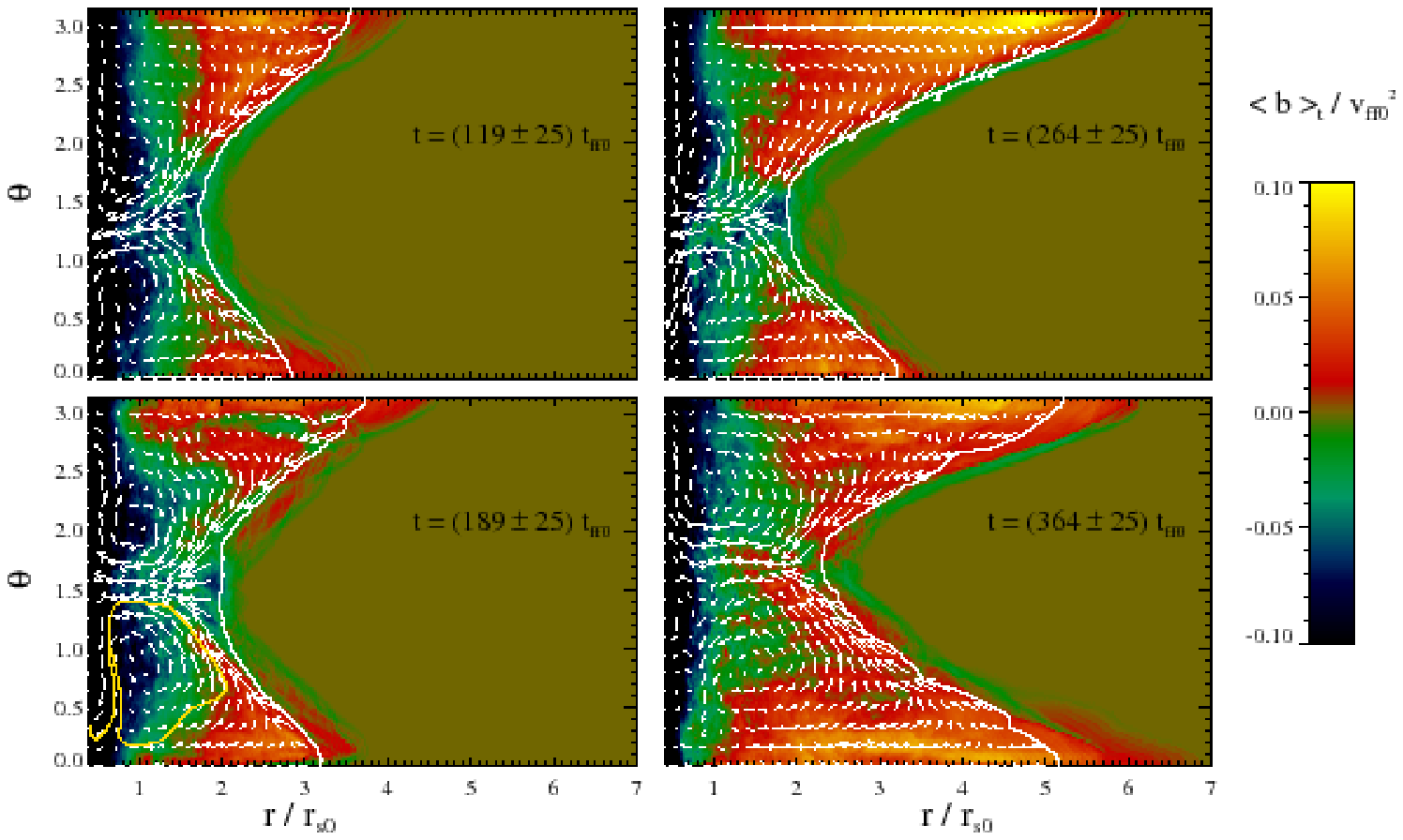}
\caption{Four snapshots of the exploding model with
$\varepsilon=0.15\vff2$ and $H = 1.02H_{\rm cr}$.  The Bernoulli
parameter (color map)
and the velocity field (white arrows)
are averaged over intervals of duration $50\,t_{\rm ff\,0}$.
The thick white contours show the surface with 50\% mass fraction in 
heavy nuclei (time-averaged). 
The yellow curve in the lower-left panel shows the result of
integrating a streamline of this time-averaged velocity field, starting 
from a point just above the radius of maximum heating. The curve performs an
overturn after $\simeq 45\tff$, and takes an extra $\simeq 8\tff$ to 
reach the inner boundary.
}
\label{f:berc_explode}
\end{figure*}

\begin{figure}
\includegraphics*[width=\columnwidth]{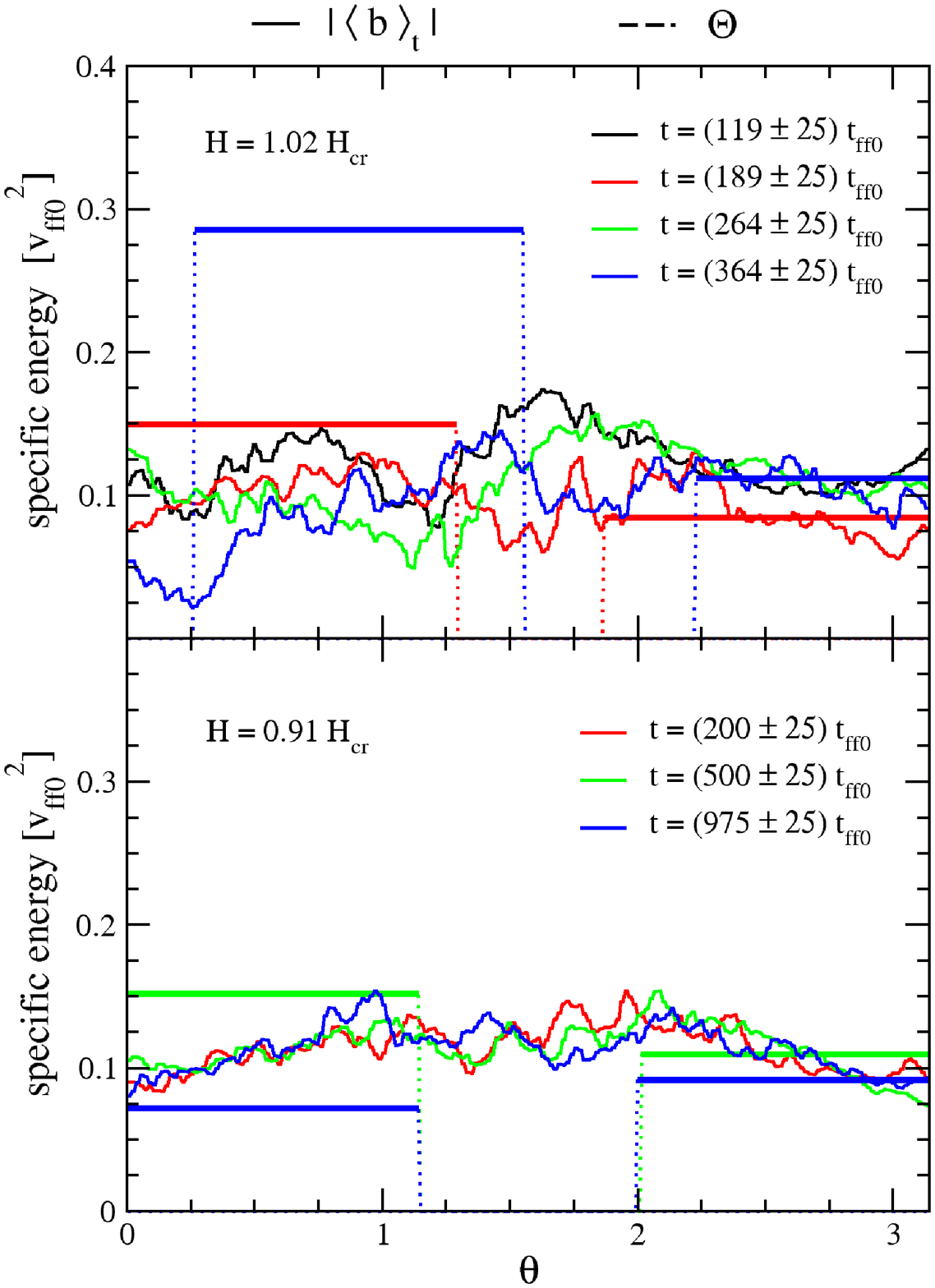}
\caption{Bernoulli parameter (thin solid lines) and specific energy absorbed during lateral advection $\Theta$ 
(equation~[\ref{eq:lateral_heat}], 
thick
lines) at the radius of maximum heating $r_{\rm H,max}$.
These
quantities are averaged over intervals of duration $50\,t_{\rm ff\,0}$.  The upper panel shows the
exploding run with $\varepsilon=0.15\vff2$ and $H = 1.02H_{\rm cr}$ (the same as in
Figure~\ref{f:berc_explode}), and the lower panel shows the non-exploding run with
$\varepsilon = 0.15\vff2$ and $H=0.91H_{\rm cr}$.
Dotted lines show the angular boundaries of convective cells. 
}
\label{f:lateral_time}
\end{figure}

\subsection{Residency Time}
\label{s:residency}

A long residency time of material in the gain region 
is commonly viewed as a key ingredient in a successful neutrino-driven 
explosion.
To calculate $t_{\rm res}$, we use the method described in \S\ref{s:setup}:
 we first assign a unique ``fluid time" $t_F$ (equation [\ref{eq:t_f}]) to each infalling
radial mass shell in the simulation, which is effectively the time
at which it passes the shock. We then define\footnote{Since $t_F$ is defined in terms of
the angle-averaged radius of the shock, there is a modest
error in $t_F$ due to non-radial deformations of the shock.
Given the lack of 
substantial large-scale mixing between the single accretion funnel and convective cells, this
prescription serves well as a tracer of different fluid populations.}
the residency time of the fluid as 
\begin{equation}
t_{\rm res} = t - t_F,
\end{equation} 
where $t$ is the present time.  A related method (tracer particles) is used by
\citet{murphy08} to calculate the residency time in collapse simulations
with a more realistic EOS.

As material with positive Bernoulli parameter accumulates below the shock, we indeed
find that its $t_{\rm res}$ grows larger.
The shock starts running outward if the energy of this unbound 
material grows on a timescale shorter than the
convective time. The right panel of Figure~\ref{f:rshock_threshold_2D} shows 
the characteristic expansion time of the shock
$t_{\rm exp}\sim r_{\rm s,max}/|\totd r_{\rm s,max}/\totd t|$
(as measured at its 
outermost radius),
alongside $\langle t_{\rm res}\rangle_{\rm vol}$ 
(as measured within the material comprising the upper part
of the residency time distribution).
The final breakout of the shock seen in the left panel of Figure~\ref{f:rshock_threshold_2D} 
corresponds to the time
when $t_{\rm exp}\sim \langle t_{\rm res}\rangle_{\rm vol}$.
This lengthening of the mean residency time can largely be ascribed to the increased
dynamical time of the expanding shock.
What changes most dramatically during breakout is the {\it ratio}
of the expansion time to the dynamical time.

The breakout is a bit more gradual in the $\rs0 = 125$ km model with heating just above
threshold for an explosion
($H = 1.04 H_{\rm cr}$; see animated version of Figure~\ref{f:energies_r125}a,b in the
online material). In this case, the expansion time of the shock remains somewhat longer
than the residency time of material below the shock, which implies that the breakout
depends on the continuing release of nuclear binding energy. 

Note that large changes in the distribution of $t_{\rm res}$ 
are concentrated in regions of positive $b$.  
In 
the left panel of
Figure\ref{f:histogram}, we plot the distribution of
$b$ and $t_{\rm res}$ 
in the $\varepsilon = 0.15\vff2$ run that is just above the threshold
for an explosion. 
Regions with small or negative residency time represent freshly injected
fluid. The distribution is stratified in $b$ around $t_{\rm res}\sim 40\tff$
(which is approximately 
an overturn period of a convective cell, see \S\ref{s:dynamics_2d}
and Figure~\ref{f:berc_explode}).
Material with more negative $b$ resides on average at a smaller radius.
Fluid  with a longer residency time has mostly positive $b$, corresponding 
to material transported upwards by convective cells.

It is also apparent from 
the right panel of
Figure~\ref{f:histogram} that
material with a longer residency time tends to have 
{\it lower} $X_\alpha$, as is expected because
it also tends to have a higher temperature.

\subsection{Heat Engine in a Two-Dimensional Explosion}
\label{s:dynamics_2d}

Fresh material that is accreted through an oblique shock has a relatively low entropy, but once it
reaches the base of the gain region it is exposed to an intense flux of electron-type neutrinos.
Some of this heated material rises buoyantly, and forces
an overturn of the fluid below the shock.  Material with a longer
residency time may therefore undergo multiple episodes of heating.
On this basis, 
\citet{herant92, herant94} suggested that a convective flow would mediate a heat
engine below the shock that would drive a secular increase in the energy of the shocked fluid.

We now investigate whether a heat engine operates in our simulations, and how it depends
on the heating parameter $H$.  We focus on a model with a constant nuclear dissociation energy,
$\varepsilon = 0.15\vff2$.  In this class of models, the infalling heavy nuclei are completely
broken up below the shock, and no heating by the reassembly of $\alpha$-particles is allowed.
As a first step, we average the convective flow
over windows of width $50\,t_{\rm ff\,0}$, which 
de-emphasizes short term fluctuations 
in the averaged velocity field $\langle {\bf v}\rangle_t$.
Figure~\ref{f:berc_explode} shows $\langle {\bf v}\rangle_t$ and $\langle b\rangle_t$
(equation [\ref{eq:bern}]) at four different times in the run
with $H$ just above the threshold for explosion ($H = 1.02H_{\rm cr}$).  
The radius of maximum heating ($r\simeq 0.66 r_\mathrm{s0}$) coincides with the lower boundary
of the convective cells, across which material flows horizontally.
The overturn period in these large scale cells is $\sim 40-50\tff$, as found by integrating
streamlines of the mean flow (an example is shown on the lower-left panel of Figure~\ref{f:berc_explode}).
Heated fluid accumulates
in the region in between the top of convective cells and the shock. 
A strong deformation of the shock allows a  plume of fresh material to descend
diagonally between the convective cells.  The tilt of this cold downflow intermittently flips in sign,
and the averaged circulation pattern typically has an ``$\infty$" shape.  The heating of fluid parcels in the
two hemispheres is also intermittent, and sometimes two circulation flows are established simultaneously, 
thereby causing a bipolar expansion of the shock.  

We have identified a useful figure of merit which connects a secular increase in the shock radius to
the strength of neutrino heating at the base of the gain region.
Figure~\ref{f:lateral_time} shows the absolute
value of the Bernoulli parameter $b$ at the radius of maximum heating 
($r_{\rm H,max}\simeq 0.66\rs0$) as a function of polar angle $\theta$.
In the top panel, the four sets of 
thin solid
lines correspond to the four snapshots of Figure~\ref{f:berc_explode},
and the bottom panel shows the analogous results for a non-exploding run. 
Overplotted as
thick solid
lines is the quantity
\begin{equation}
\label{eq:lateral_heat}
\Theta = \bigg\langle\frac{\mathscr{L}_H - \mathscr{L}_C}{\rho}\bigg\rangle_{t,\theta*,r=r_{\rm H,max}} 
\frac{r_{\rm H,max}}{|\langle v_\theta\rangle_{t,\theta*}|}.
\end{equation}
This measures the specific energy that is absorbed from neutrinos by the material
that flows laterally along the lower boundary of the convective cells. 
In equation~(\ref{eq:lateral_heat}), the angular average of the heating rate 
and meridional velocity
is restricted to a single convective cell.\footnote{To identify the 
range of angles comprising the lower boundary of a convective cell ($r = r_{\rm H,max}$), 
we first average $\langle v_\theta\rangle_t$ over all polar angles, and then define a single cell as a  
zone where $|\langle v_r\rangle_{t}|$ < $|\langle v_\theta\rangle_{t,\theta}|$ 
and $v_\theta$ maintains a constant sign.
Once the convective cells have been so identified, the angular average is repeated within each cell.
The quantity $|\langle v_\theta\rangle_{t,\theta*}|$ appearing in equation (\ref{eq:lateral_heat}) represents
this more restricted average, which typically covers
$\sim 1$ rad in the polar direction (e.g., Figure~\ref{f:berc_explode}).} 

In non-exploding models, the circulation in the gain region settles to a quasi-steady state, with no net amplification
of the mass in material with positive $b$.  The heat absorbed at the base
of the convective cells is of the same order of the Bernoulli parameter of the fluid, that is, $\Theta \lesssim |b|$.
In an exploding model, $\Theta$ will often exceed $|b|$ by a factor 2-3.
As is shown in the upper panel of Figure~\ref{f:lateral_time},
$\Theta$ grows with time as the system approaches the explosion.

The stability of the averaged flow pattern appears to be, in part, 
an artifact of the
axisymmetry of the flow.  This imposes strong restrictions on the motion of convective cells,
causing vorticity to accumulate on the largest spatial scales.
Our observation that the bulk of the neutrino heating takes place within horizontal flows
suggests that the ratio of heating timescale to radial advection time 
in the gain layer may be a less precise diagnostic of the conditions for explosion
in two dimensions: the horizontal convective velocity is typically low compared with the downward velocity
of the main accretion plume.
We do observe that the main accretion plume becomes strongly distorted near the threshold for an
explosion, so that a significant fraction of the
plume material enters one of the convection cells.  This effectively decreases the amount of material that 
accretes to the protoneutron star and thus increases the overall advection timescale across the gain region.

\section{Critical Heating Rate for Explosion}
\label{s:thresholds}

An explosion occurs when the heating parameter $H$ is raised above a critical value\footnote{Our method
for determining $H_{\rm cr}$ is discussed in \S\ref{s:1d1}.} $H_{\rm cr}$. 
We now explore how $H_{\rm cr}$ depends on the details of the EOS and the initial
radius of the shock.  One can 
express $H$ simply in terms of
the ratio of the heating rate 
($4\pi r^3 {\mathscr L}_H$) to the accretion luminosity ($GM\dot M/r$),
in the idealized (but unrealistic) case where the flow is composed only of free nucleons and
moves hypersonically.  Then this ratio depends on $H$ but not on the accretion rate
$\dot M = 4\pi r^2 \rho(r) |v_{\rm ff}(r)|$.  The precise value of the
reference radius is unimportant; we choose $r_{s0}$, the shock
radius in the time-independent, spherical flow solution at $H=0$.
Then
\begin{equation}
\label{eq:heating_ratio}
\frac{4\pi\rs0^3\mathscr{L}_H[\rho_1(r_{s0})]}
{GM\dot M/r_{s0} }{\Bigg |}_{X_\alpha = 0;\;{\cal M}=\infty} \;= \;
{2H\over \rs0 v_{\rm ff\,0}^3}.
\end{equation}
This quantity is $\sim 10^{-3}-10^{-2}$
in the models we examine, which are below or near the threshold 
for explosion.
\begin{figure}
\includegraphics*[width=\columnwidth]{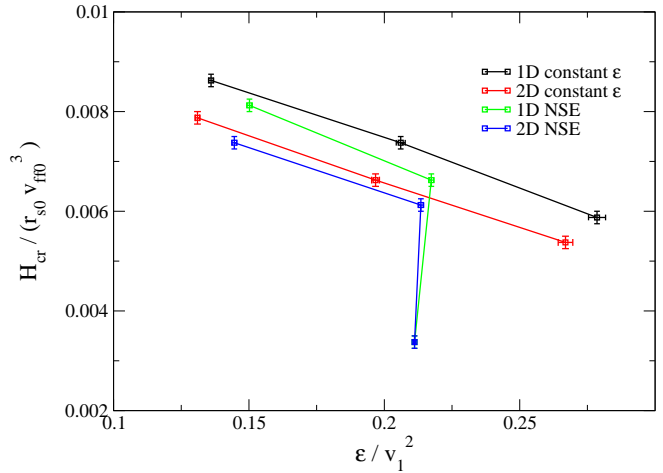}
\caption{Critical heating parameter $H_{\rm cr}$ that yields an
explosion, for all the model sequences explored in this paper
(Table~\ref{t:setups}). The abscissa is the ratio of $\varepsilon$
to $v_1^2$ in the initial flow configuration ($v_1$ being the
radial flow velocity just upstream of the shock).  Error bars
show the separation between exploding and non-exploding models,
with the points marking the average.
}
\label{f:thresholds_B_epsilon}
\end{figure}

Note that the cooling in our model is concentrated at the base of 
the settling flow.
As a result, the width of the gain region (relative to the shock radius) does
not change significantly between different models. The critical heating 
parameter
is therefore only indirectly related to the amplitude of the cooling function
through the structure of the settling flow below the shock.  Our purpose
here is to explore how the critical heating rate depends on the 
strength of the gravitational binding of the shocked fluid to the collapsed 
core, and on the abundance of alpha particles.

Figure~\ref{f:thresholds_B_epsilon} displays $H_{\rm cr}$ for all of our model sequences.
The abscissa is $\varepsilon/v_1^2$, where $\varepsilon$ is the nuclear dissociation energy 
and $v_1$ is the flow speed upstream of the shock in the initial 
configuration (that is, in the time-independent, spherical flow solution).
In the case of the NSE equation of state, this
quantity can be translated into an initial value of the shock radius
using Figure~\ref{f:Xalpha_epsilon}.  
(Note that $\varepsilon/v_1^2$ has a weak dependence on $r_s$
in the NSE sequence.)

A few interesting features of Figure~\ref{f:thresholds_B_epsilon} deserve
comment.  First, a comparison with Table~\ref{t:setups} shows that 
the critical heating rate for explosion is $\sim 50-70\%$ of the maximum
heating rate for which a steady-state flow solution can be found.
The maximal heating parameter $H_{\rm steady}$ for a steady flow
corresponds directly to
the one first determined by \citet{burrows93} using a more realistic EOS.
Note also that the values of $H_{\rm cr}$ in the one-
and two-dimensional models are much closer to each other than they are to $H_{\rm steady}$.  This result is 
perhaps not surprising, given that the explosion is not immediate,
but is approached through a series of transient fluid motions.

Second, $H_{\rm cr}$ is lower when NSE between $n$, $p$ and $\alpha$
is maintained below the shock. 
In this case, the dissociation energy at the shock is not
fixed, but is (roughly) inversely proportional to radius.
However, the difference between the NSE models
and the constant-$\varepsilon$ models is only $\sim 10\%$ in $H_{\rm cr}$ 
when the shock starts out well below $r_\alpha$ (equation [\ref{eq:r_alpha}]).
An explosion is significantly
easier when the fluid below the shock starts out with a significant population of $\alpha$ particles,
as in the models with $\rs0 = 125$~km.

\begin{figure}
\includegraphics*[width=\columnwidth]{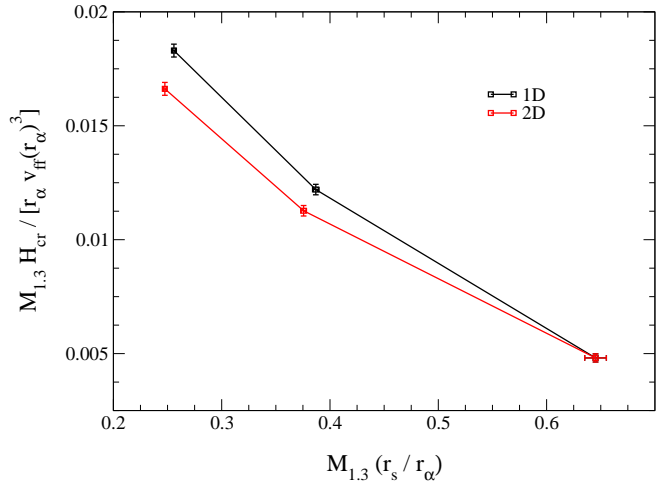}
\caption{Critical heating parameter $H_{\rm cr}$ that yields an explosion,
for the runs that include $\alpha$-particles in the EOS.  The abscissa is the
ratio of the initial shock radius $r_s$ to $r_\alpha$.
Error bars have the same meaning as in Figure~\ref{f:thresholds_B_epsilon}.
The critical heating parameter (a close analog of $L_\nu$) decreases
substantially with increasing shock radius.  The differences in
$H_{\rm cr}$ between the one- and two-dimensional models also decreases.
}
\label{f:thresholds_B_saha}
\end{figure}

Third, $H_{\rm cr}$ tends to decrease with increasing $\varepsilon/v_1^2$:
a slightly {\it lower} heating rate per unit mass
is required to explode a flow with a {\it larger} density contrast $\kappa$ across the shock.
Because almost all the gravitating mass is in the collapsed core,
the gravitational binding energy of the gain region 
is approximately proportional to $\kappa$, whereas the net heat
absorbed over the advection time is a stronger function of density,
$t_{\rm adv} \int (\mathscr{L}_H-\mathscr{L}_C) \totd^3 r \propto \kappa^2$.  (One factor of $\kappa$ comes
from the advection time $t_{\rm adv}$ as given by equation [\ref{eq:tadv}],
and the other from the density dependence of ${\mathscr L}_H$.)
For example, Table~\ref{t:setups} shows that $\kappa$ is $\sim 1.6$ times larger for $\varepsilon/\vff2=0.2$ than 
for $\varepsilon/\vff2 = 0.1$, and that $H_{\rm cr}$ is smaller by the inverse of the same factor.

Fourth, the two-dimensional runs all require less heating than their spherically symmetric counterparts to explode.  A major reason for this is that all two-dimensional
configurations explode along one or both poles 
(see Figs.~\ref{f:ephot_wad} and~\ref{f:berc_explode}),
so that less material must be lifted through the gravitational field than in a fully spherical explosion.
We have found that the precise value of 
the difference between the critical heating rate in the one- and two-dimensional
explosions depends on the choice of $r_*/\rs0$,  and therefore on the normalization $C$
of the cooling function.  The fact that we find a smaller difference than \citet{murphy08} 
may be a consequence of our simpler cooling function and equation of state.

The critical heating rate depends in an interesting way on the starting radius of the shock, in a way that points
to the recombination of $\alpha$-particles as an important last step in the transition to an explosion.
Figure~\ref{f:thresholds_B_saha} shows that $H_{\rm cr}$ in the NSE models
grows rapidly as the initial shock radius\footnote{Note that $r_s$ is the shock radius in the time-independent flow 
solution.  For a fixed cooling function, $r_s$ is a monotonically increasing function of $H$, and equals $r_{s0}$ at $H=0$.}
$r_s$ is pushed inside $r_\alpha$.  Here we normalize the heating parameter at a fixed {\it physical} radius, namely $r_\alpha$.
{\it Translated into the context of a realistic core collapse, this means that the critical neutrino luminosity for
an explosion decreases with increasing shock radius.}   The radius of the stalled shock depends, in turn, on the EOS 
above nuclear matter density:  \citet{marek07} find that a softer EOS corresponds to a larger shock radius, mainly due
to the higher accretion luminosity onto the neutronized core.  Here we have subsumed this uncertainty in the high-density EOS
into a single free parameter, the ratio $r_{s0}/r_\alpha$.  
Hydrodynamic instabilities are effective at driving an explosion 
to the extent that they push the shock radius close to $r_\alpha$; beyond
this point, the remainder of the work on the flow is done largely
by $\alpha$-particle recombination.

One also notices from Figure~\ref{f:thresholds_B_saha} that 
the difference between $H_{\rm cr}$ in one and two dimensions depends on the starting radius of the shock.
The closer $r_{s0}$ is to $r_\alpha$, the weaker the dependence of the critical heating rate
on the dimensionality of the flow.

\begin{figure}
\includegraphics*[width=\columnwidth]{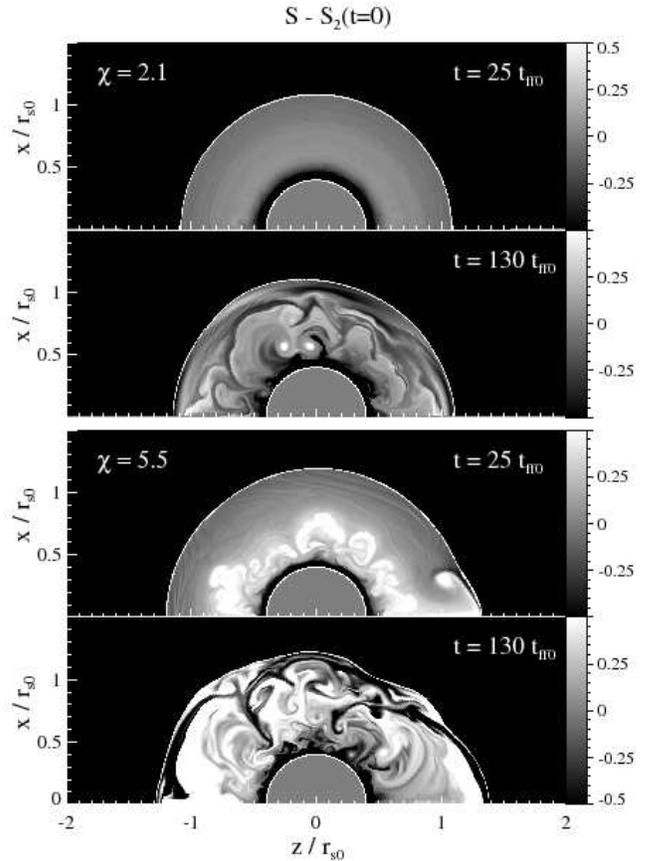}
\caption{The development of a convective instability is strongly limited
when the parameter $\chi \la 3$.  These panels show snapshots of 
entropy (normalized to initial postshock value) in
a NSE run ($r_{s0} = 75$ km) with two different heating rates.
When $\chi = 2.1$, convective cells of a limited extent are triggered
in the layer where the net heating rate is strongest, but they do not
propagate into the upper parts of the gain region.  Convection becomes
much more vigorous and widespread when $\chi = 5.5$.  Note that both
of these models are non-exploding.
An animation showing the time evolution of these two configurations
is available in the online version of the article.
}
\label{f:sasi_convection}
\end{figure}

\section{Convection and the SASI}
\label{s:turbulence}

Overturns of the fluid below the shock can be triggered in two distinct ways:  
through the development of Ledoux
convection in the presence of a strong negative entropy gradient, or via
the non-linear development of the SASI (a linear
feedback between ingoing entropy and vortex waves, and an outgoing sound wave).
We now show that the 
amplitude of the dipolar mode that is excited in the shock is 
strongly tied to the level of neutrino heating, and so 
thermal forcing plays a crucial role in maintaining the oscillation.
To a certain extent, this distinction is of secondary importance, in the sense
that memory of the linear phase of the instability is lost once
the inflow of fresh material below the shock bifurcates from older shocked
fluid.  
Nonetheless, the origin of the convective motions does have
implications
for the stability of $\ell=1$ and 2 modes in three-dimensional simulations:  one expects
that large scale oscillations will change shape and direction more rapidly
if they are triggered primarily by neutrino heating.

\begin{figure}
\includegraphics*[width=\columnwidth]{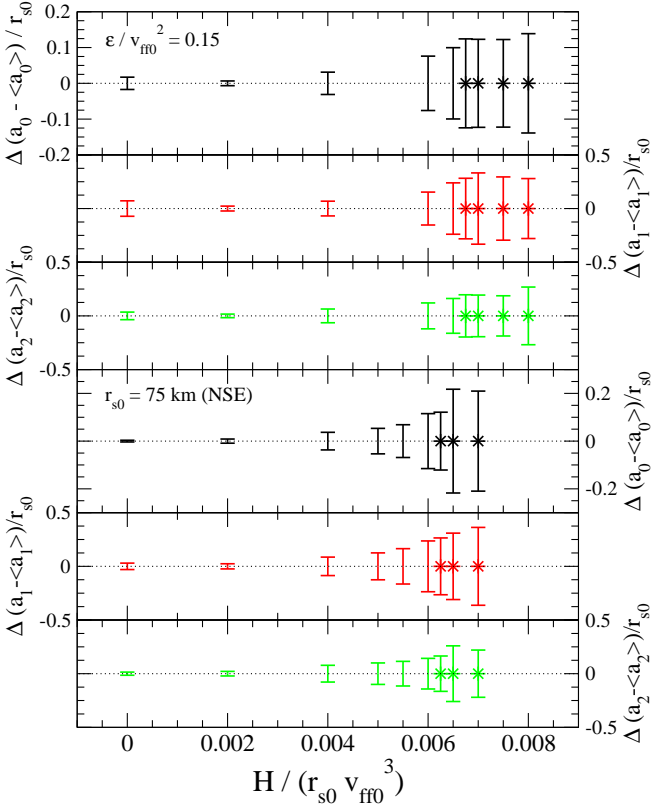}
\caption{Amplitude (r.m.s.) of $\ell = 0,1,2$ modes of the
shock in two model sequences with varying heating parameter $H$. Stars indicate exploding runs.
We show the r.m.s. fluctuation
of the difference between the instantaneous Legendre coefficient $a_\ell$ and
a running average $\langle a_\ell\rangle_{50t}$ that is computed over a window of width $50\tff$ (see text).
This subtracts
the secular movement of the shock in runs that are close to or above the threshold for explosion.
Note that the amplitude is measured in absolute units ($\rs0$).
}
\label{f:saturation_amplitudes}
\end{figure}

\begin{figure}
\includegraphics*[width=\columnwidth]{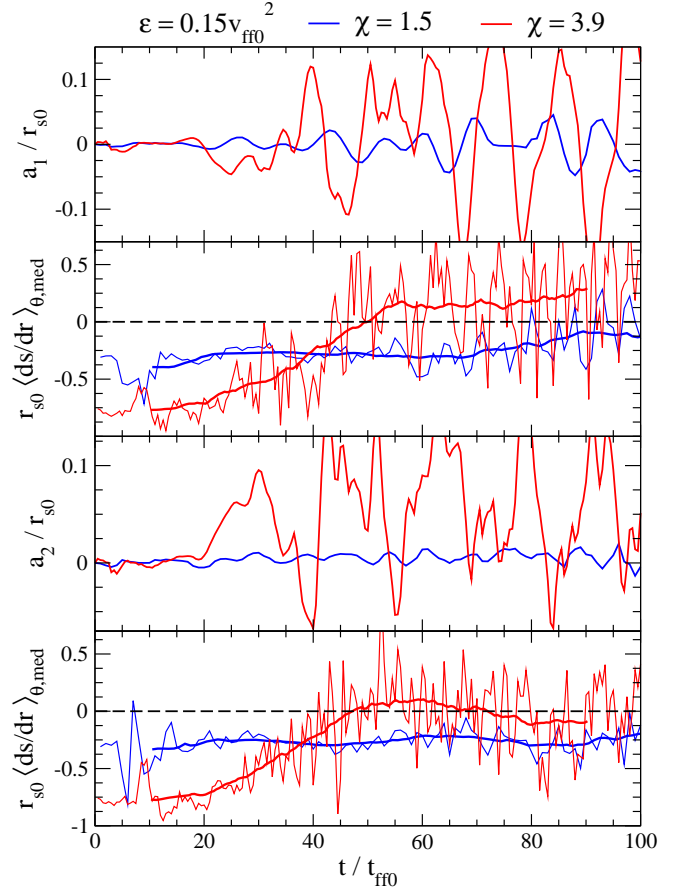}
\caption{Principal shock Legendre coefficient and
the radial median of the angle-averaged entropy gradient for
the $\varepsilon = 0.15\vff2$ model and
two different heating rates, corresponding to
$\chi = 1.5$ ($H = 0.002\,r_{s0}\,v_{\rm ff\,0}^3$, blue curves) and $\chi = 3.9$
($H = 0.004\,r_{s0}\,v_{\rm ff\,0}^3$, red curves). Both
runs are below the threshold for an explosion, but vigorous convection is established
throughout the gain region in the run with the higher heating rate.
Top (bottom) two panels:
seed perturbation is a shell with $\ell=1$ ($\ell=2$) density profile.
A running average of the entropy gradient (temporal width $20\tff$)
appears as thick solid lines.
}
\label{f:sasi_convection_e0.15}
\end{figure}

\begin{figure}
\includegraphics*[width=\columnwidth]{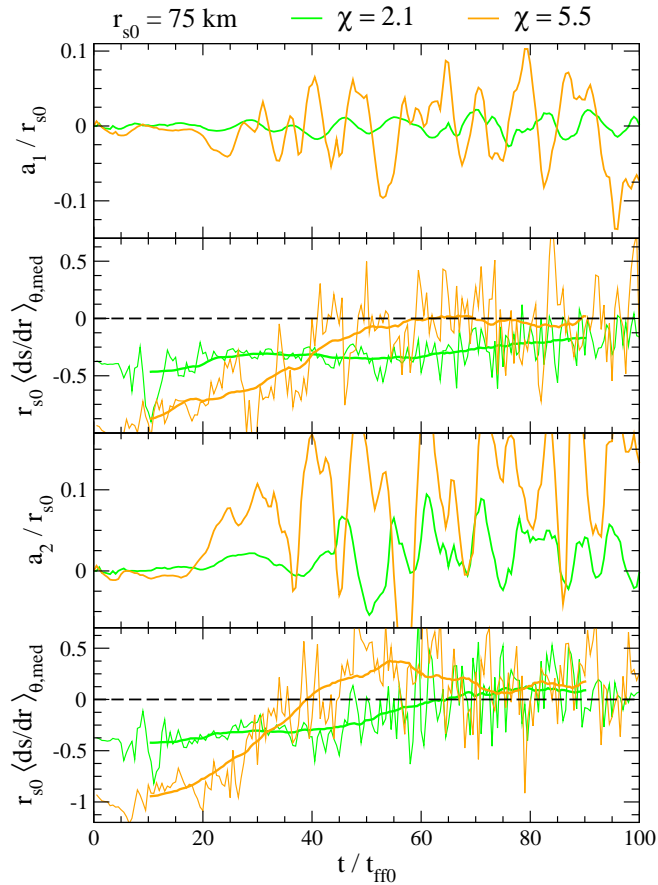}
\caption{Same as Figure~\ref{f:sasi_convection_e0.15}, but for NSE models
with $\alpha$-particles and $\rs0 = 75$~km.
}
\label{f:sasi_convection_R75}
\end{figure}

We can ask whether a heating parameter $H$ that yields an explosion will
also form an unstable entropy gradient below the shock.   
Convection develops through a 
competition between inward advection and neutrino heating.
A detailed analysis by \citet{foglizzo06} shows that the parameter
\begin{equation}
\chi \equiv \int \left|{\omega_{\rm BV}\over v_r}\right|\, dr,
\end{equation}
must exceed a critical value $\simeq 3$ for unstable convective plumes
to grow before being advected downward through
the gain region below the shock.
Here $\omega_{\rm BV}$ is the Brunt-V\"ais\"al\"a frequency.

Using our initial flow models, we can translate $H$ into a value for
$\chi$, and find (Table~\ref{t:setups}) that typically $\chi \sim 5-20$
at the threshold for a neutrino-driven explosion.  The implication
for convection below the shock is illustrated in Figure~\ref{f:sasi_convection},
which shows two snapshots for models with $\chi = 2.1$ and $5.5$,
neither of which explodes.
At the lower heating rate, the time required for convection to develop
depends on the strength of the seed perturbation, whereas at the
higher heating rate convective overturns develop rapidly within the
layer of strong neutrino heating and spread throughout the post-shock
region over a few dozen dynamical times.  (The figure shows the
result in the case where the seed perturbation is dominated by 
a small spherical startup error in the initial model.)

We conclude that, near the threshold for a neutrino-driven explosion
and for our given set of physical assumptions,
convection is driven primarily through
the development of a strong, negative entropy gradient within the gain region, rather than
through the non-linear development of SASI modes.  
The growth of the SASI requires at least a few oscillations,
each with a period comparable to the advection time.  The SASI
is therefore subdominant when $ \chi \gtrsim 3 $.  It is worth comparing
this with the results of \citet{scheck08}, who forced the inner boundary
of the simulation volume to move inward to model core contraction, thereby generating large
advection velocities.  The net result was that the flow barely reached
$ \chi \simeq 3 $ in exploding configurations.  While this effect may be
important for relatively prompt explosions, it should be kept in mind
that the rate of contraction of the neutrinosphere has slowed subtantially
a few hundred milliseconds after core bounce. \citet{marek07} found evidence 
for the delayed explosion of a $15 M_\sun$ progenitor
around 600 ms after core bounce, for which the effect of core contraction
is not likely to dominate the dynamics.

In Paper I we considered the non-linear, saturated state of the SASI in the absence of neutrino heating, 
and showed that the amplitude of the shock oscillations drops significantly as the dissociation energy 
$\varepsilon$ is increased.  We now explore how the r.m.s. amplitude of the shock oscillations correlates
with the strength of heating. 
To eliminate the effect of secular shock motions around or above the threshold for explosion, 
we first calculate the running average $\langle a_\ell\rangle_{50t}$ of the 
shock Legendre coefficients $a_\ell$ over an interval $50\tff$, and then calculate the
r.m.s. of $\hat a_\ell \equiv a_\ell - \langle a_\ell\rangle_{50t}$ 
over the duration of each simulation.
The result is plotted in Figure~\ref{f:saturation_amplitudes} as a function of $H$ for two model sequences
($\varepsilon=0.15\vff2$ and NSE with $\rs0=75$~km). 

There is a clear trend of increasing $\ell=1$, 2 mode amplitude with increasing heating rate.  This confirms
our previous suggestion that large-amplitude shock oscillations require strong heating
when the dissociation energy behind the shock exceeds $\sim 0.15\vff2$.  
The models with $H=0$ reveal a slight exception to the overall trend:  
the r.m.s. amplitude of the shock oscillations
appears larger than it does in models with small
but finite heating rate, because the oscillations are strongly intermittent
at $H=0$ (see Paper I).
The shock oscillations grow much stronger just below the threshold for explosion (exploding runs are marked by stars), above which they
seem to saturate.  Note also that their amplitudes do not vary much with the choice of dissociation model.
The r.m.s. amplitudes relative to the running average of $a_0$ at the threshold for explosion are $\{5\%, 12\%, 8\%\}$
for the $\ell={1,2,3}$ modes in the $\varepsilon/\vff2=0.15$ sequence, and $\{6\%,12\%,7\%\}$ in the NSE $\rs0=75$~km sequence.

\begin{figure}
\begin{center}
\includegraphics*[width=0.8\columnwidth]{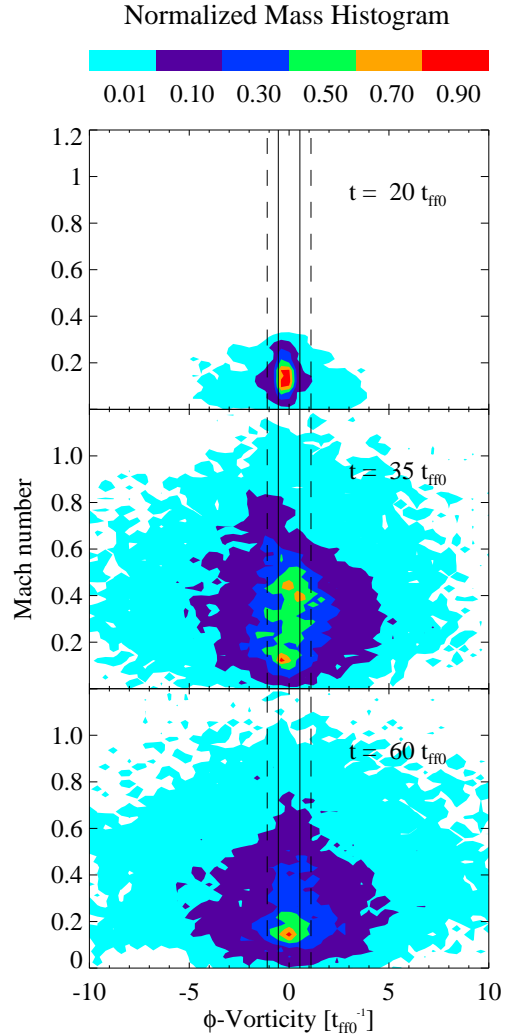}
\end{center}
\caption{
Histogram of vorticity vs. Mach number in the gain region, 
weighted by mass.
Shown are three different instants in the evolution of the run with $\varepsilon=0.15\vff2$, $\ell=1$ seed
perturbation, and $\chi = 3.9$ (corresponding to the upper two panels in Figure~\ref{f:sasi_convection_e0.15}).
The distribution broadens once 
convection is fully developed, but before the 
dipolar shock mode shows significant growth.
The vertical lines show 
the vorticity of a convective flow with period equal to (solid) 
the mean radial advection time
and (dashed) the period of a lateral sound wave at the midpoint between $r_*$ and $r_s$.  
An animated version of this figure is available in the online version
of the article.
}
\label{f:mach_vrtz}
\end{figure}

We have performed an additional sequence of runs in which
we drop an overdense shell with a given Legendre index $\ell$ 
through the shock (see also Paper I).
This has the effect of selectively triggering individual SASI modes.  
Figures~\ref{f:sasi_convection_e0.15} and \ref{f:sasi_convection_R75}
display the Legendre coefficients 
of the shock alongside the angle-averaged entropy gradient. 
We find that the amplitude of the $\ell=1$ and 2 modes is strongly tied
to the strength of convection.
For both types of dissociation models,
convection is quenched by the accretion flow when $\chi < 3$:  it grows intermittently in strength, but never reaches large enough
amplitudes to significantly distort the shock surface. As a consequence, the entropy gradient remains shallow and negative most 
of the time. 
Coherent shock oscillations are also seen, but they have a
low amplitude due to the large dissociation energy.

Convection grows much more rapidly when $\chi > 3$, 
distorting the shock surface before the SASI has the chance to execute a few oscillations.
In this case, the entropy gradient is initially more negative, but quickly flattens.
Indeed, the $\ell =1,2$ amplitudes 
only become large in models where neutrino-driven convection is strong enough to flatten
the entropy gradient.

Another way of seeing that convection is the forcing agent behind shock oscillations when $\chi > 3$
is to analyze the distribution of vorticity in the gain region. Figure~\ref{f:mach_vrtz} shows
a histogram of vorticity vs. 
Mach number  
at three different instants in the evolution of the
run with $\varepsilon = 0.15\vff2$, $\ell=1$ seed perturbation, and $\chi = 3.9$ (upper two panels
in Figure~\ref{f:sasi_convection_e0.15}). At $t=20\tff$, convection is just getting started and the 
vortical motions are restricted to Mach numbers $\la 0.3$.   
However, by $t=35\tff$ the Mach number distribution extends up to 
${\cal M} \gtrsim 0.5$ 
and has almost reached its
asymptotic form ($t=60\tff$), at 
the same time that convection has filled the region below the shock.
The dipolar mode of the shock develops a large amplitude
only after this fully developed convective state has been reached.
The convective rolls are a source of acoustic radiation (e.g. \citealt{goldreich88}), which will drive
a dipolar oscillation of the shock if the overturn frequency is comparable to the frequency of
the $\ell = 1$ mode, $|\nabla\times{\bf v}| \sim 2\times 2\pi/t_{\rm adv}$.  This zone 
is marked by the vertical solid lines in Figure~\ref{f:mach_vrtz}, and indeed encompasses most of the mass.


\section{Summary}\label{s:summary}

We have investigated the 
effects of $\alpha$-particle recombination
and neutrino heating on the hydrodynamics of core-collapse supernovae,
when the heating rate is pushed high enough to reach the threshold
for an explosion.  The effect of dimensionality has been probed
by comparing one- and two-dimensional time-dependent
hydrodynamic calculations.
Our main results can be summarized as follows:

\vskip .1in
\noindent 1. --  The critical heating parameter 
that yields an explosion depends sensitively on the starting
position of the shock relative to $r_\alpha$.   
This means that
the critical neutrino luminosity depends sensitively on the
stall radius of the shock and, in turn, on the core structure
of the progenitor star and the density profile in the forming
neutron star.  
Within the framework explored in this paper, we find two extreme
types of explosion.  In the first, neutrino heating does most of
the work, with a significant final boost from $\alpha$-particle recombination.
In the second, neutrino heating is generally less important at promoting
material below the shock to positive energies. 

\noindent 2. --  During the final stages of an explosion, 
the heat released by $\alpha$-particle recombination is comparable to
the work done by adiabatic expansion.  This heat is concentrated 
in material that has previously been heated by neutrinos.  
Significantly more energy is lost through
$\alpha$-particle dissociation in fresh downflows, so that
nuclear dissociation remains on balance an energy 
sink within the accretion flow.

\noindent 3. --
 The large-amplitude oscillations that are seen in one-dimensional runs near an explosion
are the consequence of the $\ell=0$ SASI as modified by heating.  In contrast with 
the $\ell = 1,2$ modes of a laminar accretion flow,
the period of these oscillations is close to twice the
post-shock advection time. 
The critical heating rate for an explosion (assuming
constant mass accretion rate, neutrino luminosity, and inner boundary) corresponds
to neutral stability for the $\ell = 0$ mode.

\noindent 4. 
-- The critical heating parameter $H_{\rm cr}$ for an explosion
is generally lower in two dimensions than in one, but the difference becomes smaller as
the starting radius of the shock approaches $r_\alpha$. 
The difference depends somewhat on the ratio $r_*/\rs0$ 
and thus on the cooling efficiency and equation of state.  

\noindent 5. --
Non-spherical deformations of the shock are tied to the formation of large-scale plumes
of material with positive energy.  Our
two-dimensional explosions with a super-critical heating rate
involve a large-scale convective instability that
relies on the accumulation of vorticity on the largest spatial scales.  Volume-filling
convective cells are apparent in a time-averaged sense. 
Transient heating events create positive-energy material that accumulates in between
the convective cells and the shock.
A significant fraction of the heating occurs in horizontal flows at the
base of the convective cells, which are fed by a dominant equatorial
accretion plume.
If the heating
parameter is large enough, this results in an amplifying cycle and explosion.

\noindent 6. -- 
The amplitude of the $\ell = 1$ and 2 modes correlates strongly with the value of the
heating parameter, and is coupled to the appearance of vigorous neutrino-driven convection
below the shock. In agreement with the work of \citet{foglizzo06} and \citet{scheck08}, 
we find that $\chi\approx 3$ marks the transition
from a strong linear instability in a nearly laminar flow below the shock, to
a volume-filling convective instability.  In all of our simulations,
the threshold for explosion lies well within the latter regime.
This highlights a basic difference between one- and two-dimensional explosions:
the mechanism is fundamentally non-linear in two dimensions.

\noindent 7. -- We have explored essentially  one ratio 
of cooling radius to shock radius, namely $r_*/r_{\rm s0} = 0.4$ at
zero heating (corresponding to 
$r_*/r_{\rm s} \sim 0.2$ 
near the threshold 
for an explosion).  The growth of the $\ell = 1$ SASI mode is strongest
for this particular aspect ratio when $\varepsilon = 0$ (see Figure 12 of
Paper I).  
As dissociation is introduced into the flow, we found
that the peak growth rate moves to larger values of $r_*/r_{\rm s0}$.
On the other hand, detailed collapse calculations indicate ratios
of neutrinosphere radius to shock radius that are even smaller than
$\sim 0.2$ following $\sim 100$ ms after collapse (e.g. \citealt{marek07}).
We conclude that the $l=1$ SASI mode is not being artificially suppressed
by our choice of initial shock size.

\noindent 8. -- 
Vortical motions with a Mach number
$\sim 0.3$-0.5 first appear at the onset of convective instability around the radius
of maximal neutrino heating, but before the dipolar mode 
of the shock reaches its 
limiting amplitude.  These vortices are a source of acoustic waves, which have a similar
period to the large-scale oscillation of the shock.
Near the threshold for an explosion,
the turbulence in the gain region becomes supersonic, as the existence of widespread
secondary shocks attests. These shocks convert turbulent kinetic energy to internal energy, 
increasing the effective heating rate.   

\vskip .1in
There are at least two reasons why explosions by the mechanism investigated here
may be more difficult in fully three-dimensional simulations.  
First, the existence of more degrees of freedom for the low-order modes of the
shock in three dimensions implies that the amplitude of individual shock oscillations is lower.
As a result, it is more difficult for the shock to extend out to the radius where
$\alpha$-particle recombination gives it the final push.  Second, an axisymmetric explosion that
is driven by neutrino heating involves the accumulation of vorticity on the largest spatial scales,
an effect that is special to two dimensions.  A full resolution of these issues is possible
only with high-resolution three-dimensional simulations.


\acknowledgments
We are grateful to Jonathan Dursi for scientific discussions and help with FLASH.
We also thank Adam Burrows, Christian Ott, and Jeremiah Murphy for stimulating
discussions. Careful and constructive comments by an anonymous referee helped
to improve the presentation of this paper.
The software used in this work was in part developed by the DOE-supported ASC / Alliance 
Center for Astrophysical Thermonuclear Flashes at the University of Chicago.
Computations were performed at the CITA Sunnyvale cluster, which was funded by the 
Canada Foundation for Innovation. 
This research was supported by NSERC of Canada. 
R.~F. is supported in part by the Ontario Graduate Scholarship Program.

\appendix

\section{A. Alpha-Particle Abundance in Nuclear Statistical Equilibrium}
\label{s:saha_detail}

We calculate the $\alpha$-particle mass fraction $X_\alpha$ in nuclear statistical equilibrium by
limiting the nuclear species to $\alpha$-particles and free nucleons, and fixing the electron
fraction $Y_e = 0.5$.  We tabulate $X_\alpha$ and temperature $T$ as a function of pressure $p$ and density $\rho$,
and then use these tables to calculate the rate of release of nuclear binding energy by the method
described in \S \ref{s:Saha}.  The temperature does not appear explicitly in the FLASH 
hydrodynamic
solver, and only enters the flow equations indirectly through $X_\alpha$. 

We include the contributions to $p$ from
radiation, relativistic and partially degenerate electron-positron pairs, and
nonrelativistic $\alpha$-particles and nucleons. When $k_B T>m_e c^2/2$, it can be written \citep{bethe80}:
\begin{equation}
\label{eq:total_pressure}
p = \frac{1}{12}\frac{(k_B T)^4}{(\hbar c)^3}\left[ \frac{11\pi^2}{15} + 2\eta^2 + \frac{1}{\pi^2}\eta^4\right]
    + \left(1 - \frac{3}{4}X_\alpha\right) \frac{\rho}{m_u} k_B T,
\end{equation}
where $\eta = \mu_e/(k_B T)$ the normalized electron chemical potential, 
also known as degeneracy parameter, and $\hbar$, $c$, and $m_u$ are Planck's constant, the
speed of light, and the atomic mass unit, respectively.
The density and degeneracy parameter are further related by
\begin{equation}
\label{eq:dens_eta}
\rho = \frac{m_u}{3\pi^2 Y_e}\left(\frac{k_B T}{\hbar c}\right)^3\eta(\pi^2+\eta^2),
\end{equation}   
where $Y_e$ is the electron fraction. The equilibrium fraction of $\alpha$-particles is given by the 
nuclear Saha equation,
\begin{equation}
\label{eq:saha_alpha}
X_n^2 X_p^2 = \frac{1}{2} X_\alpha \left[\frac{m_u n_Q(T)}{\rho}\right]^3 \exp{\left(-\frac{Q_\alpha}{k_B T} \right)};  \qquad
n_Q(T) = \left(\frac{m_u k_B T}{2\pi \hbar^2} \right)^{3/2}
\end{equation}
as supplemented by the conditions of mass and charge conservation,
\begin{eqnarray}
\label{eq:mass_conservation}
X_n + X_p + X_\alpha & = & 1\\
\label{eq:charge_conservation}
X_p + \frac{1}{2}X_\alpha & = & Y_e.
\end{eqnarray}
In eqs.~(\ref{eq:saha_alpha})-(\ref{eq:charge_conservation}), $X_n$ and $X_p$ are the mass fractions 
of free neutrons and protons, respectively, and $Q_\alpha = 28.3$~MeV 
is the binding energy of an $\alpha$-particle.
Combining eqs.~(\ref{eq:total_pressure}) and (\ref{eq:dens_eta}) 
gives $\eta$ and $T$ in terms of $p$ and $\rho$.  The equilibrium 
mass fraction $\Xeq$ is calculated from  $\rho$ and $T$.
For numerical calculations, we tabulate $\Xeq$, 
$\partial \Xeq/ \partial \ln{\rho}$, and $\partial\Xeq/\partial \ln{p}$
for a grid of density and pressure. In addition, we tabulate 
partial derivatives of $T$ to substitute in
eqs.~(\ref{eq:Ep_full}) and (\ref{eq:Erho_full}).

\begin{figure*}
\epsscale{1.1}
\includegraphics*[width=\textwidth]{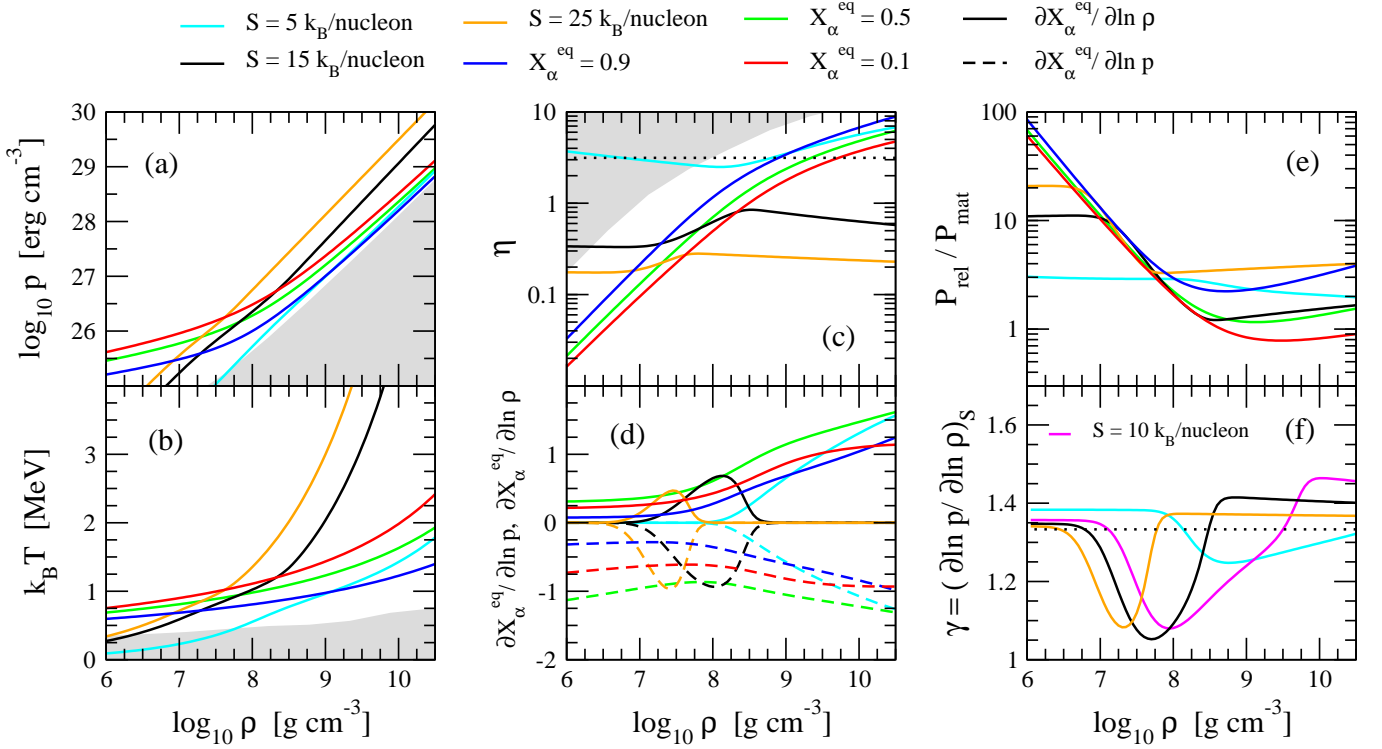}
\caption{
Equation of state of a fluid containing $n$, $p$, $\alpha$, photons, and finite-temperature 
and partially degenerate electrons, in nuclear statistical equilibrium.  We solve
eqns.~(\ref{eq:total_pressure})-(\ref{eq:charge_conservation}) and tabulate all quantities on a 
grid of density and pressure. Panels (a), (b) and (c):  pressure, temperature, and degeneracy parameter as a 
function of density, for fixed values of $\Xeq$ and entropy.   Dotted line:  $\eta = \pi$, the approximate
boundary between degenerate and non-degenerate electrons. The gray area shows the region of the thermodynamic
plane where $X_{\rm O}^{\rm eq} > 0.5$.
Panel (d):  partial derivatives of $\Xeq$ with respect to
density (solid lines, positive) and pressure (dashed lines, negative). Panel (e):  ratio of relativistic 
pressure (photons and pairs) to material
pressure ($\alpha$-particles and nucleons).  Panel (f):  adiabatic
index $\gamma$ for different adiabats (dotted line: $\gamma = 4/3$).
}
\label{f:nse_alpha_paper}
\end{figure*}

Figure~\ref{f:nse_alpha_paper} shows contours of constant $\Xeq$ and constant entropy for different variables 
as a function of density. The entropy per nucleon is obtained by adding the contributions from the different components (e.g. \citealt{bethe80}), 
\begin{equation}
\label{eq:entropy_saha}
S = \pi^2 Y_e \frac{(11\pi^2/15 + \eta^2)}{\eta(\pi^2+\eta^2)} + \left(1-\frac{3}{4}X_\alpha\right)\left[\frac{5}{2} 
    + \ln{\left\{ \frac{m_u n_Q(T)}{\rho} \right\}}\right] - X_p\ln{X_p} - X_n\ln{X_n}-\frac{1}{4}X_\alpha 
    \ln{\left(X_\alpha/32\right)}.
\end{equation}
The postshock density in the initial configuration is typically $\rho_2\sim 10^9$~g~cm${}^{-3}$ with an entropy $\sim 10-15k_B$/nucleon.
The formation of $\alpha$-particles that is seen in Figure~\ref{f:Xalpha_epsilon} results from an expansion of the shock into 
the part of the thermodynamic plane in  Figs.~\ref{f:nse_alpha_paper}a,b where $\rho_2 < 10^9$~g~cm$^{-3}$ and $T \lesssim 1$~MeV.  
In this regime, the electrons are non-degenerate and the pressure in photons and pairs begins to exceed the nucleon pressure.
The dip in the adiabatic index seen in Figure~\ref{f:nse_alpha_paper}f results from $\alpha$-particle dissociation/recombination,
which partially compensates the change in internal energy due to compression/expansion.

\section{B. Time-Independent Flow Equations Describing Initial Models}
\label{s:ode_initial}

We write down the ordinary differential equations that are used to
compute the initial flow, and the density profiles in Figure~\ref{f:profiles}a,b,c.
The steady state Euler equations in spherical symmetry are
\begin{eqnarray}
\label{eq:mass_steady}
\frac{1}{v_r}\frac{\totd v_r}{\totd r} + \frac{1}{\rho}\frac{\totd \rho}{\totd r} + \frac{2}{r} & = & 0\\
\label{eq:euler_steady}
v_r\frac{\totd v_r}{\totd r} + \frac{1}{\rho}\frac{\totd p}{\totd r} + g & = & 0\\
\label{eq:energy_steady}
\rho v_r \frac{\totd e_\mathrm{int}}{\totd r} - \frac{p v_r}{\rho} \frac{\totd \rho}{\totd r} & = & 
                                           \mathscr{L}_H - \mathscr{L}_C + \mathscr{L}_\alpha,
\end{eqnarray}
where $e_\mathrm{int}$ is the internal energy per unit mass, $\mathscr{L}_H$, $\mathscr{L}_C$, and
$\mathscr{L}_\alpha$ the source terms described in eqns.~(\ref{eq:cooling_function})-(\ref{eq:alpha_advection}),
and $g = GM/r^2$. Since two variables suffice to describe the thermodynamic state of a system, we write
\begin{equation}
\label{eq:eint_split}
\frac{\totd e_\mathrm{int}}{\totd r} \equiv E_p \frac{\totd p}{\totd r} + E_\rho \frac{\totd \rho}{\totd r}.
\end{equation}
and
\begin{eqnarray}
\label{eq:Lalpha_split}
\mathscr{L}_\alpha \equiv A_p \frac{\totd p}{\totd r} + A_\rho \frac{\totd \rho}{\totd r}.
\end{eqnarray}
The coefficients $E_i$ and $A_i$ encode the dependence on the equation of state.
Replacing eqns.~(\ref{eq:eint_split}) and (\ref{eq:Lalpha_split}) in (\ref{eq:energy_steady}), and using
eqns.~(\ref{eq:mass_steady}) and (\ref{eq:euler_steady}) to eliminate the pressure derivative, we obtain
\begin{equation}
\frac{\totd \rho}{\totd r} = \frac{\left(\rho v_r E_p - A_p \right)\left(\rho g - 2\rho v_r^2/r\right) + 
                     \left(\mathscr{L}_H -\mathscr{L}_C\right)}{\left(\rho v_r E_\rho - p v_r/\rho - A_\rho\right) 
                     + v_r^2\left(\rho v_r E_p - A_p\right)}.
\end{equation}

The coefficients in eqs. (\ref{eq:eint_split}) and
(\ref{eq:Lalpha_split}) work out to
\begin{eqnarray}
E_p    & = & \frac{1}{(\gamma-1)\rho}\qquad\qquad\quad({\rm constant}~\gamma)\\
E_\rho & = & -\frac{p}{(\gamma-1)\rho^2}\qquad\qquad({\rm constant}~\gamma).
\end{eqnarray}
For a constant-$\gamma$ equation of state, $e_\mathrm{int} = 
p/[(\gamma-1)\rho]$.  The pressure (equation~[\ref{eq:total_pressure}]) 
in the NSE model described in Appendix \ref{s:saha_detail} can be 
decomposed into contributions from relativistic particles and from nucleons,
$p = p_\mathrm{rel} + p_\mathrm{mat}$, and the 
specific internal energy is 
\begin{equation}
e_\mathrm{int} = \frac{1}{\rho}\left(3p_\mathrm{rel} + \frac{3}{2}p_\mathrm{mat} \right)
               = 3\frac{p}{\rho} - \frac{3}{2}\left( 1 - \frac{3}{4}X_\alpha \right)\frac{k_B T}{m_u}.
\end{equation}
One therefore finds
\begin{eqnarray}
\label{eq:Ep_full}
E_p    & = & \frac{3}{\rho} + \frac{9}{8}\frac{k_B T}{m_u}\frac{\partial X_\alpha}{\partial p} -
          \frac{3}{2}\left(1 - \frac{3}{4}X_\alpha\right)\frac{1}{m_u}\frac{\partial (k_B T)}{\partial p},
\qquad\qquad\quad(\textrm{NSE})\\
\label{eq:Erho_full}
E_\rho & = & -\frac{3p}{\rho^2} + \frac{9}{8}\frac{k_B T}{m_u}\frac{\partial X_\alpha}{\partial \rho}
             - \frac{3}{2}\left(1-\frac{3}{4}X_\alpha\right)\frac{1}{m_u}\frac{\partial (k_B T)}{\partial \rho}.
\qquad\qquad(\textrm{NSE})
\end{eqnarray}

The initial postshock solution is obtained by integrating the above equations
from $r_{\rm s}$ to an inner radius $r_*$ at which the flow stagnates. 
We iterate the normalization of the cooling function in equation~(\ref{eq:cooling_function})
so that $r_* = 0.4\rs0$ in the absence of heating.  When adding heating, the cooling 
normalization and $r_*$ are kept fixed, which results in an expansion of the shock from its initial position
to $r_s > \rs0$ (Figure~\ref{f:profiles}a,b). 
The initial Mach number at the inner boundary is chosen so as to satisfy
\begin{equation}
\label{s:cooling_normalization}
\bigg| \sum_i \left( \mathscr{L}_{H,\,i} - \mathscr{L}_{C,\,i} + \mathscr{L}_{\alpha,\,i}\right) V_i 
\bigg| \simeq 0.995\left[ \frac{GM}{r_*} - \varepsilon(t=0) \right] \big| \dot{M} \big|,
\end{equation}
where the sum is taken over the computational cells below the shock at our fixed resolution 
(see \S\ref{s:setup}), $V_i$ is the volume of each computational cell, and the source terms are
evaluated at the inner radial cell face. The numerical coefficient on the right hand side depends on
the radial resolution, and is chosen empirically to prevent runaway cooling due to discreteness effects
in time-dependent calculations.  
The resulting inner Mach number is typically $10^{-3}-10^{-2}$.

When including $\alpha$-particles in the EOS, one needs to calculate
self-consistently the value of $\Xeq$ below the shock, the
corresponding 
dissociation energy $\varepsilon(t=0)$ [equation~(\ref{eq:epsilon_initial})], 
and compression factor $\kappa$ [equation (\ref{eq:kappa_phot})].
The density upstream of the shock is obtained from
\begin{equation}
\label{eq:rho_1}
\rho_1(r_s) = \frac{\dot{M}}{4\pi r_s^2 |v_1(r_s)|},
\end{equation}
where
\begin{equation}
\label{eq:v_1}
v_1(r_s) = -\frac{v_\mathrm{ff}(r_s)}{\sqrt{1 + 2\mathcal{M}_1^{-2}(r_s)/(\gamma-1)}}
\end{equation}
is the upstream velocity at $r = r_s$, while the upstream pressure satisfies
\begin{equation}
\label{eq:p_1}
p_1(r_s) = \frac{\rho_1(r_s) [v_1(r_s)]^2}{\gamma \mathcal{M}_1^2(r_s)}.
\end{equation}
Eqns.~(\ref{eq:rho_1}) and~(\ref{eq:p_1}) are transformed to physical units for input to the NSE model
by adopting ${\cal M}_1(\rs0) = 5$, $\dot M = 0.3\,M_\odot$ s$^{-1}$, $M = 1.3\,M_\odot$,
and a particular value for the shock radius $r_{s0}$ in the absence of heating.

\section{C. Time Evolution}
\label{s:time_dep}

Here we give some further details of the time evolution of our initial models using FLASH2.5.
Heating and cooling are applied in an operator split way in between hydrodynamic sweeps.  Nuclear 
dissociation in the constant-$\varepsilon$ model is implemented through the 
\emph{fuel+ash}
module in FLASH, with the modifications described in Paper I. 
The rate of change of specific internal energy is computed using the current
hydrodynamic variables and timestep, after which
the EOS subroutine is called to ensure that the variables are thermodynamically consistent. 

In the NSE nuclear dissociation module, numerical stability is maintained
using an implicit update of the pressure in between hydro sweeps,
\begin{equation}
\label{eq:nse_energy_update}
p_\mathrm{new} = p_\mathrm{cur} + (\gamma-1)\rho e_\mathrm{nuc}(\rho,p_\mathrm{new}),
\end{equation}
where $e_\mathrm{nuc}$ is the energy generation per timestep 
in equation~(\ref{eq:energy_release_general}),
and the subscripts \emph{cur} and \emph{new} refer to the current and new value of the pressure, 
respectively. The density is kept constant across this step, so as to be consistent with the other
source terms. Equation~(\ref{eq:nse_energy_update}) usually converges in 3-4 Newton iterations,
adding a negligible overhead to our execution time. We restrict the timestep of the
simulation so that, in addition to the standard Courant-Friedrichs-Levy condition, it enforces 
$|e_\mathrm{nuc} | < 0.8 (p/\rho)/(\gamma-1)$. To prevent $\alpha$-particle recombination
in the cooling layer (due to the decrease of internal energy), we adopt a cutoff in density,
so that $X_\alpha = 0$ if $\rho > 3\times 10^{10}$g~cm$^{-3}$.

Both the constant-$\varepsilon$ and the NSE dissociation modules  
require that the Mach number remain below a fixed value $\mathcal{M}_{\rm burn}$ for burning, 
which has the effect of preventing dissociation or recombination upstream of the shock (Paper I). 
This results in a small amount of incomplete burning in the presence
of strong shock deformations, a phenomenon which is also
encountered in the full collapse problem. We set the threshold to
$\mathcal{M}_{\rm burn} = 2$ in most of our simulations.
The critical heating parameter $H_{\rm cr}$ depends weakly on ${\cal M}_{\rm burn}$:
changes in ${\cal M}_{\rm burn}$ cause small changes in the amount of unburnt material with zero Bernoulli parameter,
and only slightly alters the net energy of the gain region. 
At the outer boundary of the simulation volume, $\mathcal{M}_{\rm burn}$ is just
below
the Mach number of the upstream flow.  
We have tried expanded the outer boundary to $r=9\rs0$ (with a
somewhat smaller $\mathcal{M}_{\rm burn}$) and found that runs that did hit $r=7\rs0$ 
still hit the new outer boundary.

Aside from the inclusion of heating and NSE dissociation, the 
numerical setup for our runs
is identical to that of Paper I.  In that paper, the numerical output was verified by comparing the measured
growth rates of linearly unstable modes of the shock with the solution to the eigenvalue problem.
We have tested the implementation of our NSE dissociation model by verifying that, in the absence of 
initial perturbations, our steady state initial conditions remain steady.  
Spherical transients present in the initial
data die out in a few $\ell=0$ oscillation cycles, and are present even when nuclear burning is omitted.
(See Paper I for a more extended description.)

\bibliographystyle{apj}
\bibliography{references,apj-jour}

\end{document}